\documentstyle[11pt,epsf]{article}
\newcommand\Kmbf[1]{\mbox{\boldmath{$#1$}}}

\begin{document}

\newcommand{\BQ}{\begin{equation}}
\newcommand{\EQ}{\end{equation}}
\newcommand{\BQA}{\begin{eqnarray}}
\newcommand{\EQA}{\end{eqnarray}}
\newcommand{\be}{\begin{eqnarray}}
\newcommand{\ee}{\end{eqnarray}}
\newcommand{\NN}{\nonumber \\}
\newcommand{\del}{\partial}
\newcommand{\ket}[1]{\left.\left\vert #1 \right. \right\rangle}
\newcommand{\bra}[1]{\left\langle\left. #1 \right\vert\right.}
\newcommand{\ketrm}[1]{\vert {\rm #1} \rangle}  
\newcommand{\brarm}[1]{\langle {\rm #1} \vert}  

\newcommand{\kk}{k_\perp}
\newcommand{\p}{\mbox{\boldmath{$p$}}}
\newcommand{\PP}{\mbox{\boldmath{$P$}}}
\newcommand{\q}{\mbox{\boldmath{$q$}}}
\newcommand{\Q}{\mbox{\boldmath{$Q$}}}
\newcommand{\k}{\mbox{\boldmath{$k$}}}
\newcommand{\bl}{\mbox{\boldmath{$\l$}}}
\newcommand{\r}{\mbox{\boldmath{$r$}}}
\newcommand{\s}{\mbox{\boldmath{$s$}}}
\newcommand{\x}{\mbox{\boldmath{$x$}}}
\newcommand{\y}{\mbox{\boldmath{$y$}}}

\newcommand{\bfb}{\mbox{\boldmath{$b$}}}
\newcommand{\bfc}{\mbox{\boldmath{$c$}}}

\def\simge{\mathrel{
   \rlap{\raise 0.511ex \hbox{$>$}}{\lower 0.511ex \hbox{$\sim$}}}}
\def\simle{\mathrel{
   \rlap{\raise 0.511ex \hbox{$<$}}{\lower 0.511ex \hbox{$\sim$}}}}

\begin{flushright}
SACLAY-T04/096
\end{flushright}

\begin{center}
\vspace{3cm} 
{\LARGE\bf Light Mesons on the Light Front \\}
\vspace{1.5cm}
{\large
 K. Naito$^{\, a,}$\footnote{\tt knaito@nucl.sci.hokudai.ac.jp}, \ 
S. Maedan$^{\, b,}$\footnote{\tt maedan@tokyo-ct.ac.jp}\  and  \ 
K. Itakura$^{\, c,}$\footnote{\tt itakura@spht.saclay.cea.fr} 
}
\\
\vspace{1.3cm}

{\it $^a$ 
    Meme Media Laboratory, 
    Hokkaido University, Sapporo 060-8628, Japan}\\
{\it $^b$ Department of Physics, Tokyo National College of Technology, 
Tokyo 193-0997, Japan}\\
{\it $^c$ Service de Physique Th\'eorique, CEA/Saclay, F91191 
Gif-sur-Yvette Cedex, France}
\vspace{13mm}
\end{center}

\vspace{.3cm}

\begin{abstract}
We study the properties of light mesons in the scalar, pseudo-scalar, 
and vector channels within the light-front quantization, 
by using the (one flavor) Nambu--Jona-Lasinio model with 
vector interaction.
After taking into account the effects of chiral symmetry breaking,
we derive the bound-state equation in each channel in the 
large $N$ limit ($N$ is the number of colors), 
which means that we consider the lowest $q\bar q$ Fock state 
with the {\it constituent} quark and antiquark. By solving the 
bound-state equation, we simultaneously obtain a mass and a 
light-cone (LC) wavefunction of the meson.  
While we reproduce the previous results for the scalar and 
pseudo-scalar mesons, 
we find that, for a vector meson, the bound-state equations 
for the transverse and longitudinal polarizations 
look different from each other.
However, eventually after imposing a cutoff which is invariant under 
the parity and boost transformations, one finds these two are identical,
giving the same mass and the same (spin-independent) LC
wavefunction. When the vector interaction becomes larger than 
a critical value, the vector state forms a bound state, whose mass decreases
as the interaction becomes stronger.
While the LC 
wavefunction of the pseudo-scalar meson is broadly distributed
in longitudinal momentum ($x$) space, that of the vector meson 
is squeezed around $x=1/2$. 
\end{abstract}

\newpage
\tableofcontents

\newpage

\section{Introduction}
\setcounter{equation}{0}

Understanding the structure of light mesons at low energies is 
one of the most difficult problems in QCD, which
requires nonperturbative studies about three entangled issues: 
chiral symmetry breaking, confinement of quarks and gluons, 
and relativistic bound-state physics. 
The light-front (LF) quantization is now widely accepted as 
the most transparent formalism for the third issue, the
relativistic bound-state physics \cite{LF_review}, and thus, 
if the other two are also accommodated by the LF formalism, 
then it can be a very powerful tool for the problem. 
This expectation has been shared by many people with continuous 
interests and efforts in the LF formalism. 
The present paper is one of such attempts towards
establishing the methods for describing the chiral symmetry 
breaking on the LF. 

The advantage of the
LF formalism in treating the relativistic bound-state 
physics is closely related to the fact that the vacuum in this 
formalism is very simple, and is actually the Fock vacuum itself 
even for an interacting system. Roughly speaking, this property
allows us to construct
a Hamiltonian equation $H_{\rm LC}\ket{\alpha}=E\ket{\alpha}$
with only a few Fock states just like in the Tamm-Dancoff approximation. 
By solving this equation, one can simultaneously find the wavefunction 
of each Fock state and the eigen-value (mass) of the eigen state. 
This is certainly the unique merit of the LF formalism.
On the other hand,
the other two properties, chiral symmetry breaking and confinement,
 are usually believed to be related to the nontrivial vacuum structure. 
This does not mean that the LF quantization 
is not useful for the problem mentioned above, but rather means that 
one has to discover new ways of 
formulating the physics of confinement and chiral symmetry 
breaking within the context of the LF quantization. 
This motivated several efforts towards setting up the problems in QCD,
and actually many new interesting ideas and techniques were generated 
in such activities (see for example, Ref.~\cite{Wilson}). However, 
we have not reached at a satisfactory description for these problems. 

Meanwhile, there has been considerable progress in describing the 
spontaneous symmetry breaking in scalar theories within the 
LF quantization \cite{LF_review}. It was recognized that the 
longitudinal zero mode of a scalar field plays a crucial r\^ole. 
The scalar zero mode is not a dynamical mode in the LF quantization,
and subject to a non-linear constraint equation, called the 
zero-mode constraint.
Spontaneous symmetry breaking can be achieved through the 
nonperturbative solution
to the zero-mode constraint, which brings in a nonzero vacuum 
expectation value to the scalar field. 
Moreover, based on the analogies with such achievements, 
even the chiral symmetry breaking in simple fermionic models 
(without gauge fields) were described within the LF
quantization \cite{Itakura-Maedan_DLCQ, Itakura-Maedan_NJL} 
(See Ref.~\cite{Itakura-Maedan_review} for a review about these 
approaches based on the canonical LF quantization. 
For an alternative treatment based on the path-integral approach, 
see Ref.~\cite{Lenz}).
Therefore, the region governed by the LF formalism
is now slowly expanding, while the chiral symmetry breaking 
in real QCD is still not under control (see, however, 
Ref.~\cite{Zhang}).

Indeed, within the Nambu--Jona-Lasinio (NJL) model, 
which is sometimes considered as an effective theory of QCD 
\cite{NJL_review,Hatsuda-Kunihiro}, 
it has been found  that, 
still with the trivial Fock vacuum, chiral symmetry breaking 
is described in such a way that one selects an appropriate 
Hamiltonian depending on the phases of 
the symmetry \cite{Itakura-Maedan_NJL}. 
In the NJL model, different Hamiltonians 
are originated from different solutions to a constraint equation, 
which exists only in the LF formalism. The "bad" component of 
the spinor
is not a dynamical variable and is subject to a constraint equation
like the zero mode in scalar theories.
 This "fermionic constraint" is  nonlinear 
in the NJL model and leads to the "gap equation" for the 
chiral condensate.
When the coupling constant  is larger than the critical value, 
the gap equation has a non-zero solution even in the chiral limit.
This means that the fermionic constraint allows for "symmetric" and
"broken" solutions corresponding to those of the gap equation. 
If one selects the "broken" solution, and substituting it to the canonical 
Hamiltonian, one obtains the "broken" Hamiltonian. This 
governs the dynamics in the broken phase and is completely 
different from the Hamiltonian with the "symmetric" solution.
In Ref.~\cite{Itakura-Maedan_NJL}, two of us solved the fermionic 
constraint by using the $1/N$ expansion, 
and obtained the Hamiltonians in both symmetric
and broken phases. They also solved the bound-state equations
for the scalar and pseudo-scalar mesons and obtained their 
light-cone (LC) wavefunctions and masses, as well as the PCAC 
and Gell-Mann, Oakes, Renner (GOR) relations.

The main objective of the present paper is to extend the analysis 
of Ref.~\cite{Itakura-Maedan_NJL} to the vector channel by using 
a similar model, and investigate the properties of a vector meson. 
A part of the results was already reported 
without detailed derivations \cite{PLB}. Thus, in this paper, we will 
re-derive all the results in a self-contained way as much as 
possible. We will also present a slightly deeper understanding of the 
properties of scalar and pseudo-scalar mesons.

It will be helpful to briefly summarize the main results of the 
present paper.
After lengthy calculations, we are able to derive $q\bar q$ 
bound-state equations in the pseudo-scalar ($\pi$), scalar ($\sigma$), 
and vector (V) channels. By solving these equations, we will eventually 
obtain the LC wavefunctions and masses of the mesons. Remarkably, 
it turns out that, 
due to the contact interaction of the NJL model, 
the spin-independent part of the LC wavefunction for each meson 
($i=\pi,\, \sigma,\,$V) has a very simple form 
(see Eq.~(\ref{Solution_general})):
$$
\phi_i(x,\kk)\propto \frac{1}{m_i^2-\frac{\kk^2+M^2}{x(1-x)}},
$$
where $x$ is the fraction of longitudinal momentum carried by a quark, 
$\kk$ is the relative transverse momentum between the quark and the antiquark,
$M$ is the constituent quark mass, and $m_i$ is the mass of the meson $i$.
This result is common for all the mesons discussed in the paper. 
The difference among the LC wavefunctions is visible only through the 
value of the meson mass $m_i$. For example, in the chiral limit, we will 
find a relation $0=m_\pi<m^{}_{\rm V}<m_\sigma=2M$.
 Namely, the LC wavefunction of the 
pseudo-scalar has a broad distribution in $x$ space, 
while that of the scalar meson is
rather peaked around the mean value $x=1/2$, and the vector meson
comes in between these two (see Fig.~\ref{LCWF}).

The paper is organized as follows.
In the next section, we will define the 
NJL model with the vector 
interaction. The vector interaction is added to have a bound state 
in the vector channel.
We also show the explicit form of the fermionic constraint of 
this model.
In Sect.~\ref{FC_Ham}, we solve the fermionic constraint by using 
the $1/N$ expansion, and derive the LC Hamiltonian 
to the first nontrivial order.
This allows us to derive the bound-state equations of light mesons,
which is discussed in Sect.~\ref{Bound_State}. To the leading 
nontrivial order of the $1/N$ expansion, mesons are described by 
a quark-antiquark state, and the bound-state equations can be 
restricted to the lowest $q\bar q$ Fock states. 
In Sect.~\ref{LCWF_mass}, we solve the bound-state equations 
in scalar, pseudo-scalar, and vector channels, and obtain 
the masses of mesons and the LC wavefunctions as already discussed above. 
We will show that, when we treat the bound-state equations, 
we have to be extremely careful about the regularization 
scheme. Otherwise, the equations for transverse and longitudinal
polarizations of the vector meson look differently.
 Summary and discussions are given in Sect.~\ref{Summary}. Some detailed 
calculations, as well as the notation we use are presented in 
Appendix.

\section{The model}\label{The_model}
\setcounter{equation}{0}

In this section, we define the NJL model we use, 
and discuss its particular properties when it is quantized on 
the light front.
We also clarify our strategy which we take in the present paper.

\subsection{Nambu--Jona-Lasinio model with vector interaction}
\label{NJL}
The simplest NJL model is given by 
\begin{eqnarray}
{\cal L}_{\rm NJL} &=& \bar\Psi_a(i \partial \!\!\!/ -m_0)\Psi_a
  +{\cal L}_{\rm S}\, ,\NN
{\cal L}_{\rm S}&\equiv&\frac{G_{\rm S}}{2}
\left[(\bar\Psi_a\Psi_a)^2+(\bar\Psi_a i \gamma_5 \Psi_a)^2\right]\, ,
\label{NJL_Lag}
\end{eqnarray}
where $\Psi_a$ is a spinor in the fundamental representation 
of "color" SU($N$) group ($a$ is a color index: $a=1,\cdots,N$). 
 The scalar 
interaction ${\cal L}_{\rm S}$ is constructed as "color" singlet. 
In what follows, we suppress the color index $a$ for notational 
simplicity, which however does not cause any confusion because,
throughout the paper except this section, we always treat color 
singlet objects (i.e., bilocal operators to be defined below). 
We consider the case with one flavor for simplicity. 
Generalization to the case with multi flavors, 
which is necessary for phenomenological applications, 
is more complicated but is straightforward.
Recently, this model was analyzed in the light-front quantization 
by two of the authors \cite{Itakura-Maedan_NJL} (see also 
Ref.~\cite{Itakura-Maedan_DLCQ}) and they computed 
masses and light-cone wavefunctions of scalar and pseudo-scalar 
mesons to the first nontrivial order 
of the $1/N$ expansion. 
One can of course apply the same procedure for vector states, but 
we know that the simplest NJL model defined by 
${\cal L}_{\rm NJL}$ does not 
allow for a bound state in the vector channel \cite{NJL_review, Klimt90}. 
Vector states start to bind if one adds the vector interaction 
so that the attractive force between a quark and
an antiquark in the vector channel becomes stronger. 
Therefore, we include the vector interaction minimally by adding 
the following interaction:
\begin{equation}
{\cal L}_{\rm V} = - \frac{G_{\rm V}}{2} \left[ (: \bar{\Psi}\gamma_\mu\Psi :)^2
 + (: \bar{\Psi}\gamma_\mu\gamma_5 \Psi :)^2 \right]. \label{vector_int}
\end{equation}
Notice that this interaction does not change the nontrivial
"vacuum structure" (more precisely, the value of chiral condensate) 
which is caused by ${\cal L}_{\rm S}$. This is  
because we have taken the normal order in (\ref{vector_int}) 
which is defined with respect to the Fourier modes of the 
fermion field (see below for the details).
On the other hand, the bound-state equations for scalar and 
pseudo-scalar mesons will be modified due to the vector interaction. 

It would be very helpful to summarize here the issues to be discussed 
in the present paper and our strategy against them. 
There are mainly three issues:
(i) chiral symmetry breaking, (ii) $q\bar q$ bound states 
in the scalar and pseudo-scalar channels, and (iii) $q\bar q$ bound 
states in the vector channel. Below we briefly explain the roles 
played by the interactions ${\cal L}_{\rm S}$ and ${\cal L}_{\rm V}$ 
for these problems. The precise meaning of all 
the procedures and approximations will be clarified in the 
forthcoming sections. 
\begin{enumerate}
\item[(i)] {\bf Chiral symmetry breaking} is generated by the 
interaction ${\cal L}_{\rm S}$, and is not affected by the inclusion 
of the vector interaction
${\cal L}_{\rm V}$ (by construction) 
to the leading order of the $1/N$ expansion. 
 
\item[(ii)] {\bf Bound states in the scalar and pseudo-scalar channels}
are possible\footnote{Precisely, to the first nontrivial leading order of the 
$1/N$ expansion, the scalar meson appears as a bound state only in the 
chiral limit. Besides, even with the vector interaction included,
this result is the same because it affects only the pseudo-scalar 
channel.} 
in the presence of the scalar interaction 
${\cal L}_{\rm S}$.
Thus, in this paper, we will study them with the Lagrangian (2.1).
Although the vector interaction  ${\cal L}_{\rm V}$ will modify the
 scalar and pseudo-scalar bound states, we shall ignore the effects of
${\cal L}_{\rm V}$ expecting that the scalar interaction ${\cal
L}_{\rm S}$ gives the dominant effect.

\item[(iii)] {\bf Bound states in the vector channel}
are not formed with ${\cal L}_{\rm S}$ alone, but can exist with
the help of ${\cal L}_{\rm V}$. Thus, to study the bound-state 
physics in the vector 
channel, we need to include the effects of ${\cal L}_{\rm V}$. 
In the actual calculation, we will take the leading contribution
of the eigen-value equation with respect to $G_{\rm V}$ which is 
enough to form a bound state.

\end{enumerate}
Therefore, in the following analysis, inclusion of the vector 
interaction (\ref{vector_int}) is only relevant for the bound-state 
physics of the vector state. 

\subsection{A constraint equation for the bad spinor component}
One of the main sources of complexities in the light-front quantization
is the presence of constraint equations which appear only when we  
treat the light-cone variable $x^+$ as the evolution time. 
In a system involving a fermion field, one always 
has a constraint equation for the "bad" component of the spinor 
$\psi_-$ (where $\psi_\pm\equiv \Lambda_\pm \Psi=
\frac12 \gamma^\mp \gamma^\pm \Psi$ and 
$\psi_+$ is refereed to as "good", see Appendix A for our notations) 
because the kinetic term 
$i\bar\Psi\del\!\!\!/\Psi$ is expressed as 
$i\psi_+^\dagger \del_+\psi_++i\psi_-^\dagger \del_-\psi_-+\cdots $
and $\del_-$ is now a spatial derivative $\del_-=\del/\del x^-$.  
This constraint equation which we call the "fermionic constraint", 
is easily obtained if we multiply the Euler-Lagrange equation by 
the projector $\Lambda_+$ from the left
($\Lambda_+\gamma^-\del_-\Psi=\gamma^-\Lambda_-\del_-\Psi=
\gamma^-\del_-\psi_-$):
$$
\Lambda_+\left[ (i \partial \!\!\!/ -m_0)\Psi + \frac{\delta}{\delta \bar\Psi } ~  ({\cal L}_{\rm S}+{\cal L}_{\rm V})\right]=0\, .
$$
Without the vector interaction, the fermionic constraint takes a
rather simple form:
\begin{eqnarray}
i \partial_- \, \psi_- =  \frac{1}{2} \,
 \left( i \gamma^j_\perp  \partial_{\perp j} + m_0 \right) \gamma^+ \psi_+
  - \frac{G_{\rm S}}{2} \Big[ 
  \gamma^+ \psi_+ (\bar{\psi}_+ \psi_- + \bar{\psi}_- \psi_+ ) 
 - i\gamma_5 \gamma^+ \psi_+
 (\bar{\psi}_+ i\gamma_5 \psi_- + \bar{\psi}_- i\gamma_5 \psi_+)
 \Big].  \NN
  \label{FCwoVector}
\end{eqnarray}
While this is just a first order differential equation, 
the term coming from the interaction makes 
the analysis difficult. Still, in a classical level, this equation 
can be exactly solved \cite{Itakura-Maedan_NJL}, which helps to 
understand properties of the "chiral transformation" on the light front
that is different from the ordinary one because the chiral 
transformation is imposed only on the dynamical ("good") 
component~\cite{Itakura-Maedan_review}. 
But, as was discussed in Ref.~\cite{Itakura-Maedan_NJL}, in order to describe 
the symmetry breaking, we need to solve the fermionic constraint 
in a {\it quantum} level, which we will indeed do in the next section. 
It should be noticed that, in the quantum theory, the operator 
ordering must be specified, though it is a priori not known.
Thus in Ref.~\cite{Itakura-Maedan_NJL}, they {\it defined} the quantum theory 
by the above operator-ordering. Namely, $\psi_+$ comes to the left of 
color singlet operators such as $\bar\psi_+\psi_-$ or 
$\bar\psi_+ i \gamma_5 \psi_-$.
Since this operator ordering 
correctly reproduces physics of the chiral symmetry breaking,
we adopt the same operator ordering in the present paper.

With the vector interaction included, the corresponding fermionic 
constraint becomes terribly complicated, while its structure
(as a differential equation)
is the same as Eq.~(\ref{FCwoVector}). We show its explicit 
form for later convenience:
\begin{eqnarray}
 i\partial_- \psi_- \!\!& =&\!\! \frac{1}{2}
 \left( i\gamma^i_\perp \partial_{\perp i} + m^{}_0\right)
  \gamma^+ \psi_+ 
- \frac{G_{\rm S}}{2} \Big[ 
  \gamma^+ \psi_+ (\bar{\psi}_+ \psi_- + \bar{\psi}_- \psi_+ ) 
 - i\gamma_5 \gamma^+ \psi_+
 (\bar{\psi}_+ i\gamma_5 \psi_- + \bar{\psi}_- i\gamma_5 \psi_+)
 \Big]  
 \nonumber \\
 &&\hspace*{-5mm}+\, G_{\rm V}\, \Big[ \psi_- (:\bar{\psi}_+ \gamma^+ \psi_+:)
 + \gamma_5 \psi_- (:\bar{\psi}_+ \gamma^+ \gamma_5 \psi_+:) \Big] 
 \nonumber \\
 &&\hspace*{-5mm}+\, \frac{G_{\rm V}}{2} \left[
  \gamma^i_\perp \gamma^+ \psi_+
 :\left\{ \bar{\psi}_+ \gamma^i_\perp \psi_- +
 \bar{\psi}_- \gamma^i_\perp \psi_+ \right\}: 
 -  \gamma_\perp^i \gamma_5 \gamma^+ \psi_+
 :\left\{ \bar{\psi}_+ \gamma^i_\perp \gamma_5 \psi_- +
 \bar{\psi}_- \gamma^i_\perp \gamma_5 \psi_+ \right\}: 
 \right]. \hspace{9mm} \label{FCwithVector}    
\end{eqnarray} 
Here we have followed the same operator ordering as 
in Eq.~(\ref{FCwoVector}). This equation only makes 
sense in the quantum level because we have already 
taken the normal order.

\subsection{Quantization}
Since the bad component of the spinor is not an independent 
degree of freedom, the quantization condition is imposed only 
on the good component. From Dirac's procedure, one finds the 
following anti-commuting relations:
\begin{eqnarray}
&& \left\{\psi_{+ \alpha}(\x),\, \psi_{+ \beta}^{\dagger}(\y)\right\}_{x^+=y^+}
=\frac{1}{\sqrt{2} } ( \Lambda_+ )_{\alpha\beta} \
                  \delta(\x-\y)\, , \label{quantization1}\\
&&\Big\{\psi_{+ \alpha}(\x),\, \psi_{+ \beta}(\y)\Big\}_{x^+=y^+} 
=\left\{
\psi_{+ \alpha}^\dag (\x),\, \psi_{+ \beta}^{\dagger}(\y)\right\}_{x^+=y^+} 
=0\, , \label{quantization2}
\end{eqnarray}
where\footnote{
Notice that $\x=(x^-,x_\perp^i)$ for space coordinates, but 
$\p=(p^+,p_\perp^i)$ for momenta, and 
$p\cdot x =p^-x^+-\p\x$ with $\p\x=-p^+x^-+k_\perp^i x_\perp^i$. 
See Appendix \ref{Notation}.}
 $\delta(\x - \y)\equiv 
\delta(x^- - y^-)\, \delta^{(2)}(x_\perp -y_\perp).$
The mode expansion of the field is given by 
\begin{equation}\label{mode_expansion}
 \psi_+ (\x) =  \,
\int\frac{ [d\p]}{\left(2\pi\right)^{3/2}}\, \frac{1}{\sqrt{2 p^+}}  
   \sum_{s=\pm} \left[\, b(\p, s)\, u_+(\p, s) 
   \, {\rm e}^{-i p \cdot x}
  \, + \, d^{\dagger} (\p, s) \, v_+(\p, s)
     \, {\rm e}^{i p \cdot x}\, \right],  \nonumber 
\end{equation}
where we have introduced notation for the integration 
$$
\int [d\p]\equiv \int_0^\infty dp^+ \int_{-\infty}^\infty d^2p_\perp\, ,
$$
and $u_+(\p, s)$ and $v_+(\p, s)$ 
are the good components of the solutions (with spin $s$) 
to the free Dirac equation, i.e., 
$u_+ \equiv \Lambda_+ u$ and $v_+ \equiv \Lambda_+ v$ (for more details, 
see Appendix B).
Notice that the longitudinal momentum $p^+$ takes only positive values
because of the special property of the dispersion relation on 
the light front $p^-=(p_\perp^2+m^2)/2p^+>0$.
The operators $b( \p, s  )$ for quarks 
and $d ( \p, s  )$ for anti-quarks satisfy the 
anti-commutation relations:
\begin{eqnarray}
\left\{ b (\p,s),\, b^{ \dagger}(\p', s') \right\}
= \left\{ d (\p,s),\, d^{ \dagger}(\p', s') \right\}
= \delta_{ s s'}\, \delta(\p - \p'),
\end{eqnarray}
and zeros for other combinations. 
The normal order in Eq.~(\ref{FCwithVector}) is defined with respect to 
these operators.

\section{Fermionic constraints and LC Hamiltonians}\label{FC_Ham}
\setcounter{equation}{0}

Here we solve the fermionic constraint in the quantum theory, 
and compute the LC Hamiltonian which is one of the 
necessary ingredients for deriving the
bound-state equations. While the whole procedure, with 
the vector interaction included, is really tedious, it is 
straightforward and essentially the same as in the case without 
the vector interaction which was recently achieved in 
Ref.~\cite{Itakura-Maedan_NJL}. 
Thus, in this section, we first discuss 
the case without 
the vector interaction in order to demonstrate the methods we use. 
The presentation is basically along the previous analysis of 
Ref.~\cite{Itakura-Maedan_NJL}, but we here adopt the spinor basis 
for the mode expansion (\ref{mode_expansion}) (cf. instead of a 
simple Fourier expansion, Eq.~(3.2) in Ref.~\cite{Itakura-Maedan_NJL}) 
and do not use any specific representations of the $\gamma$ matrices, 
both of which are much more convenient for the extension to the case 
with vector interaction. 
 In order to avoid too much technicalities, but 
still to keep the paper self-contained as much as possible, we 
put some of the details aside to Appendix \ref{Hamiltonian}.
The involved case with vector 
interaction follows after this, with emphasis only on the
differences coming from the inclusion of the vector interactions.
Notice also that the analysis without the vector interaction can be
applied to the scalar and pseudo-scalar mesons since we simply ignore 
the effects of vector interaction in these channels.

\subsection{The case without vector interaction as a warm-up}
\subsubsection{Solving the fermionic constraint in bilocal form}

It was found in Ref.~\cite{Itakura-Maedan_NJL} that one can systematically 
solve the fermionic constraint by using the $1/N$ expansion 
after rewriting it with respect to color singlet bilocal operators. 
The $1/N$ expansion of the bilocal operator equations 
is made possible by use of the boson expansion method, 
which has been known as a standard technique in many body physics, 
but was only recently applied to the light-front field theories 
in Ref.~\cite{Itakura_BEM}.
The basic quantity is defined by a color singlet bilocal operator
of the good component at different space points and at the same LC time:
\BQ
{\sf M}_{\alpha \beta}(\x ,\y ) \equiv
  \sqrt{2}\, \psi_{+ \alpha}^{ \dag}(\x ) \, 
             \psi_{+ \beta } (\y )\, ,\label{bilocal_M}
\EQ
where $\x=(x^-,x_\perp^i)$ and 
$\alpha,\, \beta$ are the Dirac indices, and the color indices
are suppressed.  
We further introduce  bilocal operators of 
scalar (${\sf S}$) and pseudo-scalar
(${\sf P}$)  as follows: 
\begin{eqnarray}
      {\sf S}_{\rm R}(\x ,\y )  & \equiv & \bar \Psi (\x )
             \Lambda_- \Psi  (\y )
             \, +\, {\rm h.c.} ,  \label{def_SR}  \\
      {\sf P}_{\rm R} (\x ,\y ) & \equiv &  \bar \Psi (\x )
\Lambda_- (i \gamma_5) \Psi  (\y )
           \, + \, {\rm h.c.} ,\label{def_PR}
\end{eqnarray}
where h.c. is the hermitian conjugate and the subscript "R" 
implies that it is the "real" part 
(later we will define the "imaginary" part when we discuss 
the case with vector interaction).
Notice that  these bilocal operators reduce to the scalar and 
pseudo-scalar operators at the same space-time point:
${\sf S}_{\rm R}(\x ,\x )  = \bar \Psi (\x ) \Psi (\x )$ 
and 
${\sf P}_{\rm R}(\x ,\x )=\bar \Psi (\x ) i\gamma_5 \Psi (\x )$.
Notice also that these quantities depend upon the bad component of 
the spinor $\psi_-$ and thus are not known until we solve the fermionic
constraint. In other words, once we rewrite the fermionic constraint 
(\ref{FCwoVector}) with respect to the bilocal operators, they can be 
interpreted as the constraint equations for these unknown 
bilocal operators. In particular, the equations for ${\sf S}_{\rm R}$ 
and ${\sf P}_{\rm R}$ form a closed set of equations.
Indeed, one obtains in momentum space 
\begin{eqnarray}
q^+ {\sf S}_{\rm R} (\p,\q ) &=&
  \frac12 \left(\gamma^i q_\perp^i  +m_0\right)_{\alpha\beta}
               \Big({\sf M}_{\alpha\beta}(\p,\q ) - 
  {\sf M}_{\alpha\beta}(\q ,\p)
                  \Big)       \nonumber \\
   &&-\, \frac{G_{\rm S}}{2}\int  \frac{d^3 \k\, d^3 \bl}{(2\pi)^3}
     \left\{ {\sf M}_{\alpha\beta}(\p, \q -\k -\bl)
         \Big( {\sf S}_{\rm R} 
            + i \gamma_5 \, {\sf P}_{\rm R} 
         \Big)_{\alpha\beta}(\k ,\bl) \right.  \nonumber \\
   && \hskip2.5cm   \left.
       -\, \Big( {\sf S}_{\rm R} 
            - i \gamma_5 \, {\sf P}_{\rm R} 
         \Big)_{\alpha\beta}(\k ,\bl)\, 
              {\sf M}_{\alpha\beta}( \q -\k -\bl, \p )
                \right\} \,  , \label{FC_mom_S}\\ 
q^+ {\sf P}_{\rm R} (\p,\q ) &=&
    - \, \frac{1}{2}  \left\{  (i \gamma_5) ( \gamma^i q_\perp^i + m_0 )
\right\}_{\alpha\beta}
                  {\sf M}_{\alpha\beta}(\p,\q )
 \, - \, \frac{1}{2}  \left\{  (  \gamma^i q_\perp^i + m_0 ) (i
\gamma_5) \right\}_{\alpha\beta}
                  {\sf M}_{\alpha\beta}(\q ,\p)
\nonumber \\
   && +\,\frac{G_{\rm S}}{2}\int \frac{d^3 \k\, d^3 \bl}{(2\pi)^3}
     \left\{ {\sf M}_{\alpha\beta}(\p, \q -\k -\bl)
         \Big(  i \gamma_5 \, {\sf S}_{\rm R} 
            - {\sf P}_{\rm R}  \Big)_{\alpha\beta}(\k ,\bl)
\right.  \nonumber \\
   && \hskip2.5cm   \left.
       + \, \Big( i \gamma_5 \, {\sf S}_{\rm R}
            + {\sf P}_{\rm R}  \Big)_{\alpha\beta}(\k ,\bl)\, 
              {\sf M}_{\alpha\beta}( \q -\k -\bl, \p )
                \right\} ~ ,\label{FC_mom_P}
\end{eqnarray}
where we have used short-hand notations like 
$({\sf S}_{\rm R}(\p,\q))_{\alpha\beta}\equiv\delta_{\alpha\beta}\, 
{\sf S}_{\rm R}(\p,\q)$, $( {\sf S}_{\rm R} 
            + i \gamma_5 \, {\sf P}_{\rm R} 
         )_{\alpha\beta}(\k ,\bl) \equiv ( {\sf S}_{\rm R} (\k ,\bl)
            + i \gamma_5 \, {\sf P}_{\rm R} (\k ,\bl)
         )_{\alpha\beta},$
etc, and the momentum integration is from $-\infty$ to $\infty$
(even for longitudinal momenta), 
which follows from our definition of the 
Fourier transformation
$$
{\sf M}_{\alpha\beta}(\p,\q )\equiv 
\int_{-\infty}^\infty\frac{d^3 \x}{(2\pi)^{3/2}}
     \int_{-\infty}^\infty\frac{d^3 \y }{(2\pi)^{3/2}}\
     {\sf M}_{\alpha\beta}(\x ,\y )\ \, 
{\rm e}^{-i{\p\x}  - i {\q \y} }~.
$$
We have written equations only for ${\sf S}_{\rm R}$ and ${\sf P}_{\rm R}$ 
(instead of a generic matrix with spinor indices constructed by 
good and bad components like $\psi^\dagger_{+\alpha}\psi_{-\beta}$),
but these two equations are enough for the purpose of computing 
the Hamiltonian (see below).

We will solve Eqs.~(\ref{FC_mom_S}) and (\ref{FC_mom_P}) 
by using the $1/N$ expansion. 
We expand the bilocal operators as follows:
\begin{equation}
 {\sf M}_{\alpha\beta}(\p,\q ) = N\sum_{n=0}^\infty
\left(\frac{1}{\sqrt{N}}\right)^n
       {\sf m}_{\alpha\beta}^{(n)}(\p,\q )~, 
\label{M_expansion}
\end{equation}
and similarly for ${\sf S}(\p,\q)$ and ${\sf P}(\p,\q)$, and 
the expansion coefficients of them are 
written as ${\sf s}^{(n)}(\p,\q)$ and ${\sf p}^{(n)}(\p,\q )$, 
respectively 
(for notational simplicity, we omit the subscript R 
in this subsection).
Each contribution ${\sf m}^{(n)}(\p,\q)$ is given by the boson 
expansion method of the Holstein-Primakoff type 
\cite{Itakura_BEM,Itakura-Maedan_NJL} as a function of a bilocal 
bosonic operator $B(\p_1,s_1:\p_2,s_2)$  satisfying 
\begin{equation}
  \Big[ B(\p_1,s_1:\p_2,s_2 ),\, B^\dagger(\p'_1,s'_1 : \p'_2,s'_2 ) 
  \Big]
   = \delta_{s_1 s'_1 } \, \delta(\p_1-\p'_1)\,
     \delta_{s_2 s'_2 } \, \delta(\p_2-\p'_2)\, ,
\end{equation}
and so on.
Note that there is no $N$-dependence in ${\sf m}^{(n)}(\p,\q)$ 
since $B(\p_1,s_1:\p_2,s_2)$ is of ${\cal O}(N^0)$.
Note also that this bilocal operator is in the spinor basis 
(i.e., with quantum numbers (spin) $s_1,s_2$), and is not the same 
as the one in the Dirac basis (i.e., with the Dirac indices like 
$\alpha,\beta$) introduced in Ref.~\cite{Itakura-Maedan_NJL}. 
The spinor basis is more convenient for
treating the vector mesons than the Dirac basis.
The boson expansion method is constructed in such a way that the 
commutator which is satisfied by the bilocal operator 
${\sf M}_{\alpha\beta}(\p,\q)$ is correctly fulfilled by using the 
bosonic operator $B(\p,s,\p',s')$. It is useful to recognize that the 
commutator between two $:\!{\sf M}\!:$'s becomes {\it bosonic} in the large 
$N$ limit. Thus, in this limit, 
the bosonic operator $B$  (or $ B^\dag$) becomes identical to 
the bilocal operator $:\!{\sf M}\!:$ (depending on the sign of the momenta) 
and thus can be understood as annihilation/creation operators of 
a quark-antiquark pair.
The definition of this expansion and explicit form of 
the first few terms are given in Appendix C.

Substituting the lowest order expression (\ref{HP_lowest}) of the boson 
expansion, one can compactly rewrite the lowest order ${\cal O}(N)$
of the fermionic constraints (\ref{FC_mom_S}) and 
(\ref{FC_mom_P}) as
\begin{equation}
  \pmatrix{
    {\sf s}^{(0)}(\p, \q )  \cr
    {\sf p}^{(0)}(\p, \q )}
  = \pmatrix{
  m_0 \frac{\epsilon(p^+)}{q^+} \delta (\p+ \q )\cr
   0        }
   -g^{}_{\rm S}\frac{\epsilon(p^+)}{q^+}\int_{-\infty}^\infty
      \frac{d^3 \k  } {(2\pi)^3}
   \pmatrix{
         {\sf s}^{(0)}(\k , \p+\q -\k  )       \cr
        {\sf p}^{(0)}(\k , \p+\q -\k  )       \cr
     },\label{LOFC}
\end{equation}
where we have rescaled the coupling constant $ G_{\rm S} = g^{}_{\rm S}/ N $
(Recall that the large $N$ limit should be taken with $G_{\rm S}N$ being 
fixed). The solutions are easily obtained:
\begin{eqnarray}
  {\sf s}^{(0)}(\p, \q )  &=&  -M\frac{\epsilon(p^+)}{p^+}
                               \delta(\p+\q  )~, \label{sol_lowest} \\
  {\sf p}^{(0)}(\p, \q )  &=&  0~,
\end{eqnarray}
where $M$ is the dynamical mass of the fermion and is given 
self-consistently by
\begin{eqnarray}
 M & \equiv &  m_0 -G_{\rm S} \bra{0}  \bar\Psi\Psi \ket{0}
\nonumber \\
     & =&  m_0 -  g^{}_{\rm S} \int \frac{d^3 \p\, d^3 \q  }{(2\pi)^3}  ~
           {\sf s}^{(0)}(\p, \q )
             + {\cal O}({1}/{\sqrt{N} } )\, .
\end{eqnarray}
As we will see below, this equation becomes the gap equation after we 
impose an appropriate cutoff. Substituting the solution (\ref{sol_lowest}) 
into the above definition of $M$, one encounters an integral 
which is divergent both in infra-red and ultra-violet regimes:
\BQ
\frac{M-m_0}{M}=g^{}_{\rm S}\int \frac{d^3\p}{(2\pi)^3} 
\frac{\epsilon(p^+)}{p^+}\, .\label{gap_integral}
\EQ
We regularize this integral by introducing a cutoff
$\Lambda$. We have to admit that even physical 
quantities may have explicit dependence upon the cutoff $\Lambda$
because the NJL model is not a renormalizable theory. 
Besides, it is important to know that the cutoff cannot be 
taken arbitrary. 
The key for finding an appropriate cutoff is to keep as many 
symmetries of the system as possible. 
It turned out \cite{Itakura-Maedan_NJL} that the 
parity\footnote{The parity transformation in the $z$ direction 
corresponds to the exchange $p^+\leftrightarrow p^-$ in momentum space.} 
invariant cutoff 
$|p^\pm | <\Lambda$ yields a reasonable result (even in the chiral limit)
\begin{equation}
\frac{M-m^{}_0}{M}
=\frac{g^{}_{\rm S}\Lambda^2}{4\pi^2}\left\{2-\frac{M^2}{\Lambda^2}
\left(1+ \ln \frac{2\Lambda^2}{M^2}\right)\right\} \, ,
\label{gap}
\end{equation}
where we have used the dispersion relation of a free fermion 
 with the dynamical mass
$p^2=2p^+p^--p_\perp^2=M^2$ to derive the lower limit of the $p^+$ integral.
Imposing a cutoff on both the LC energy and the longitudinal momentum 
seems to violate the boost invariance 
of the system, but this is not the case because 
$p^\mu$ is an {\it internal} momentum to be integrated out.
We will, however, have to be more careful when we apply the similar 
cutoff to the two body sector.

\begin{figure}
 \epsfxsize=14cm
 \centerline{\epsffile{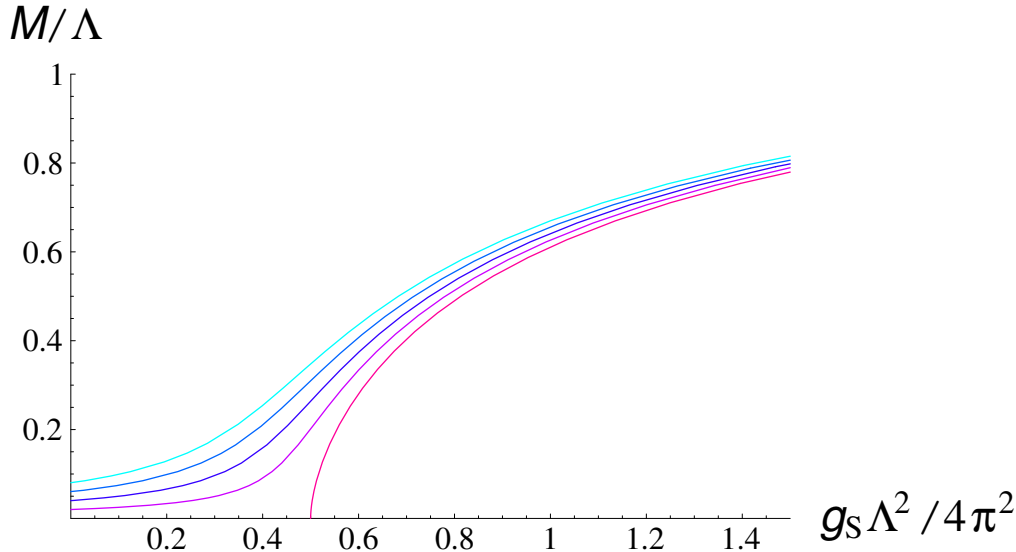}}
 \caption{Dynamical quark mass $M/\Lambda$ as a function of 
the dimensionless coupling constant 
$g_{\rm S}\Lambda^2/4\pi^2$ for different current quark masses, 
  $m_0/\Lambda=0.08,\, 0.06,\, 0.04,\, 0.02,\, 0$ 
(chiral limit), from top to bottom.}
\label{condensate}
\end{figure}

When the coupling constant $g^{}_{\rm S}$ is larger  
than the critical value $g^{\rm (crit)}_{\rm S}=
2\pi^2/\Lambda^2$,
the gap equation (\ref{gap})  develops a nonzero solution even in 
the chiral limit. This is easily confirmed by the numerical  
solution to the gap equation, as is shown in Fig.~1. Therefore,
if we take this nonzero solution\footnote{Strictly speaking, the 
symmetric solution for $g^{}_{\rm S}>g^{\rm (crit)}_{\rm S}$ 
appears only in the chiral limit. If the current
quark mass is taken non-zero and is decreased to zero, 
then the solution naturally approaches to the
broken solution for $g^{}_{\rm S}>g^{\rm (crit)}_{\rm S}$,
and to the symmetric one for $g^{}_{\rm S}<g^{\rm (crit)}_{\rm S}$.
This is evident on Fig.~\ref{condensate}.
The symmetric solution for stronger coupling is unstable against
the small explicit violation of the chiral symmetry, just like in the 
Ising model under the external magnetic field.} 
and substitute it to the canonical 
Hamiltonian, we are able to describe the dynamics in the broken phase.

As we will discuss shortly, solving the bilocal fermionic constraints 
(\ref{FC_mom_S}) and (\ref{FC_mom_P}) up to the next-to-leading order 
is important for deriving the LC Hamiltonian to the first nontrivial order. 
In fact, one can even solve the constraints at arbitrarily high order,
in the inductive way. Namely, if one knows the solutions up to the  $n$-th 
order, then the $(n+1)$-th order solution is given by some known functions and 
the solutions whose orders are lower than $n+1$. 
All the details are discussed in Appendix~\ref{Hamiltonian},
together with the derivation of the LC Hamiltonian.

\subsubsection{The LC Hamiltonian}

Since the canonical LC Hamiltonian  
is bilinear with respect to the spinor (see Eq.~(\ref{Ham_wo_Vector})), 
one can rewrite it by the bilocal operators:
\begin{eqnarray}
 H\, =\, P^-  
  & = & \frac{1}{2} \int d^3 \x  \int d^3 \y\,  \partial^y_j ~
        {\sf V} ^j_{\rm I} (\x, \y)\, \delta (\x - \y)
     +   \frac{m_0}{2} \int d^3 \x  ~{\sf S}_{\rm R} (\x, \x)  \nonumber  \\
    & = & \frac{1}{2} \int d^3 \p \int d^3 \q  ~( i q_\perp^j ) 
        {\sf V}^j_{\rm I} ( \p, \q)\, \delta ( \p + \q )
         +   \frac{m_0}{2} \int d^3 \p  ~{\sf S}_{\rm R} (  \p, -\p)\, ,
\label{Ham_bilocal}
\end{eqnarray}
where we have introduced a new bilocal operator,
\begin{eqnarray} \label{def_VI}
   i \, {\sf V} ^j_{\rm I} (\x, \y) & \equiv  &   \bar \Psi (\x) ~
     \Lambda_-  \gamma^j_\perp  \Psi (\y)
     -  {\rm h.c.}  \ \, . 
\end{eqnarray}
Note that this new bilocal operator can be related to the known operators,
${\sf M,\, S,\, P}$ (see Appendix~\ref{Hamiltonian}). 
Therefore, given the solutions ${\sf s}^{(n)}_{\rm R}$ and 
${\sf p}^{(n)}_{\rm R}$, one can 
immediately obtain ${\sf v}_{\rm I}^{i\, (n)}$, the expansion coefficients 
of ${\sf V}_{\rm I}^{i}$.
Hence, one can compute the LC Hamiltonian order by order
\begin{eqnarray}
  H &=& N\sum_{n=0}^\infty\left(\frac{1}{\sqrt{N}}\right)^n
                h^{(n)}_{\rm S}~,
\label{Ham_expand}
\end{eqnarray}
where the subscript S was put in order to remind that this is 
for the case with ${\cal L}_{\rm S}$ (\ref{NJL_Lag}) 
(without the vector interaction).
We substitute the "broken" solutions of ${\sf s}^{(n)}_{\rm R}$ and 
${\sf p}^{(n)}_{\rm R}$ to obtain the Hamiltonian which describes 
the broken phase of the chiral symmetry. It turns out that the 
lowest order $ h^{(0)}_{\rm S}$ 
is divergent but is just a constant, and thus can be neglected for the
present purpose. 
The next order ${\sf v}_{\rm I}^{ i ~ (1) }  (\p,{-\p}) $ and
${\sf s}^{ (1) }_{\rm R}  (\p,{-\p}) $ become strictly zero, and 
we have $ h^{ (1) }_{\rm S} =0 $. 
Therefore, the first nontrivial contribution is given by the 
next-to-leading order $n=2$.
After straightforward calculation, one arrives at 
the following expression (see Appendix~\ref{Hamiltonian} for the derivation):
\begin{eqnarray}
h^{(2)}_{\rm S}
\!\! &=& \!\!  
   \int [d  \p] [d  \q] ~ \left\{\frac{p_\perp^2 + M^2}{2p^+} 
                               + \frac{q_\perp^2 + M^2}{2q^+} 
                          \right\}
    \sum_{s_1,s_2} B^\dag(\p, s_1 : \q , s_2 ) B(\p, s_1 : \q , s_2 )
                    \nonumber  \\
  & &\!\!  - \, g^{}_{\rm S}\kappa^{}_{\rm S}
 \int \frac{[d  \p] [d  \q ] [d  \k ] [d  \bl]} {(2\pi)^3}     \,
     \frac{\delta(  \p  +\q -\k -  \bl) }{ 4 \sqrt{  p^+ q^+ k^+ l^+ } } \, 
                 \sum_{s_1,s_2} \sum_{s'_1,s'_2}B^\dag(\p, s_1 : \q , s_2 ) 
                          B(\k , s'_1 : \bl, s'_2 )      \nonumber  \\
  & & \!\!   \times   
          \biggl[  \Big\{ {\bar u} ( \p,s_1)  v( \q,s_2) \Big\}\,   
                   \Big\{ {\bar v} (\bl,s'_2)  u( \k,s'_1) \Big\}   
         + \Big\{ {\bar u} ( \p,s_1) \, i \gamma_5 \, v(\q,s_2) \Big\}\, 
           \Big\{ {\bar v} ( \bl,s'_2) \, i \gamma_5 \, u( \k,s'_1) \Big\}
             \biggr]\, ,\ \ \ \ \ \ 
          \label{h2_wo_Vector}
\end{eqnarray}
where we have introduced a constant
\BQ
\kappa^{}_{\rm S} \equiv \left\{ 1+ g^{}_{\rm S}\int \frac{d^3 \r}{(2\pi)^3}
\frac{\epsilon(r^+)}{P^+-r^+} \right\}^{-1}.
\label{kappa_S}
\EQ

As already mentioned, in the leading order of the $1/N$ expansion, 
the bosonic operator $B^\dag(\p,s_1:\q,s_2)$
can be interpreted as a creation operator of a quark-antiquark 
pair (the pair behaves like a boson). 
Since this Hamiltonian contains only terms of $B^\dag B$ type, it 
describes the dynamics of the two body $q\bar q$ sector.
Indeed, the first, diagonal, term corresponds to the free part for 
this quark-antiquark pair with constituent mass $M$, which 
is easily understood by the fact that 
$(p_\perp^2+M^2)/2p^++(q_\perp^2+M^2)/2q^+$ 
corresponds to the sum of LC energies of a free quark and a free 
antiquark. 
The second term, which is proportional to the (scaled) coupling 
constant and thus clearly comes from the interaction, needs to be 
diagonalized. As we will see later, diagonalization leads to 
a bound-state equation for the quark and anti-quark pair.

It is of special importance to recognize that the divergent factor 
$\kappa^{}_{\rm S}$ defined by Eq.~(\ref{kappa_S}) is independent 
of the external momentum 
$P^+$. This is formally seen by changing the variable in the 
longitudinal integration ($y=r^+/P^+$)
$$
\int_{-\infty}^\infty dr^+\frac{\epsilon(r^+)}{P^+-r^+}=
\int_{-\infty}^\infty dy\, \frac{\epsilon(y)}{1-y}\, .
$$
Of course we need to regularize this integral, but if we use a cutoff
which does not depend on the external momentum $P^+$, then the result 
does not depend on $P^+$, either. This is true as far as we treat a 
boost invariant cutoff since the external momentum $P^+=p^++q^+$
is eventually related to the total momentum of the two body $q\bar q$ state
(see discussion in the next section).

\subsection{The case with vector interaction}

Inclusion of the vector interaction certainly complicates the 
fermionic constraint, but the structure of the constraint 
(as a differential equation) is the same and 
we can follow the same procedure as before.
Thus, we do not repeat the whole procedure here, 
but rather present only the differences from the previous case.
The main difference is that we have to treat another types of 
bilocal operators, which makes the number of equations double.

\subsubsection{Definition of the bilocal fields}

In the case without the vector interaction,
the fermionic constraint (\ref{FCwoVector}) was transformed into
two equations (\ref{FC_mom_S}) and (\ref{FC_mom_P}) which form a 
closed set for bilocal operators ${\sf S}_{\rm R}$ and ${\sf P}_{\rm R}$.
On the other hand, when the vector interaction is turned on, 
the original fermionic constraint (\ref{FCwithVector}) cannot 
be expressed only with respect to 
${\sf S}_{\rm R} $ and  $ {\sf P}_{\rm R}$, but we need to introduce 
another types of operators, the vector ${\sf V}^\mu$ and 
the axial vector ${\sf A}^\mu$.
Specifically, we define the following bilocal operators:
\begin{eqnarray}
   {\sf S}(\x ,\y )  & \equiv & \bar \Psi (\x ) \Lambda_- \Psi (\y )\, ,
\qquad \qquad
 {\sf P}(\x ,\y )   \equiv  \bar \Psi (\x ) \Lambda_-   ( i
                               \gamma_5 ) \Psi  (\y ),\\
{\sf V}^i(\x ,\y )   &\equiv&  \bar \Psi (\x ) \Lambda_-   (
\gamma_\perp^i )  \Psi  (\y )\, ,\quad \quad 
  {\sf V}^\pm(\x ,\y )   \equiv  \frac12 \bar \Psi (\x )
\gamma^\pm  \Psi  (\y ),\\
 {\sf A}^i(\x ,\y )   &\equiv&  \bar \Psi (\x ) \Lambda_-   (
 \gamma_\perp^i  \gamma_5 )  \Psi  (\y )\, ,\quad
  {\sf A}^\pm(\x ,\y )   \equiv  \frac12 \bar \Psi (\x )
\gamma^\pm   \gamma_5 \,  \Psi  (\y ).
\end{eqnarray}
For each operator ${\sf O=S,P,V^\mu,A^\mu}$, 
we further define its "real" and "imaginary" 
parts as follows:
\begin{eqnarray}
  {\sf O}(\x ,\y ) &=&  \frac{1}{2}\, \Big\{ {\sf O}_{\rm R}(\x,\y )
                      + i \, {\sf O}_{\rm I}(\x ,\y ) \Big\}\, ,
\end{eqnarray}
where 
\begin{eqnarray}
  {\sf O}_{\rm R}(\x ,\y ) &=&  {\sf O}(\x ,\y ) + 
 {\sf O}(\x ,\y )^\dagger,  \nonumber \\
  i \, {\sf O}_{\rm I}(\x ,\y ) &=&  {\sf O}(\x ,\y ) -
{\sf O}(\x ,\y )^\dagger,    \nonumber 
\end{eqnarray}
so that they satisfy 
${\sf O}_{\rm R,I}(\x ,\y )^\dagger =  {\sf O}_{\rm R,I}(\x,\y )$.
Of course these definitions are consistent with  (\ref{def_SR}), 
(\ref{def_PR}) and (\ref{def_VI}), which were already introduced 
when we  solved the fermionic constraint
without vector interaction. Note that all the "real" parts 
at the same space-time points (i.e., ${\sf O}_{\rm R}(\x,\x)$) 
reduce to the  operators 
which have appropriate statistics. Namely,
\begin{eqnarray}
 && {\sf S}_{\rm R}(\x , \x ) = \bar \Psi (\x ) \Psi (\x),\quad \,
 \quad\ {\sf P}_{\rm R}(\x,\x )= \bar \Psi (\x)\, i\gamma_5\, \Psi (\x), 
 \label{Interpolating1}\\
 && {\sf V}_{\rm R}^\mu(\x ,\x ) 
            = \bar \Psi (\x )\, \gamma^\mu\, \Psi (\x ),\quad
  {\sf A}_{\rm R}^\mu(\x ,\x ) 
            = \bar \Psi (\x )\, \gamma^\mu  \gamma_5 \, \Psi  (\x ).
\label{Interpolating2}
\end{eqnarray}
Actually, the operator ${\sf O}(\x,\y)$ has been constructed so that
${\sf O}_{\rm R}(\x,\x)$ becomes the standard bilinear operator with 
appropriate statistics. These "interpolating" operators 
become useful later when we 
determine the kinematical structure of the lowest Fock state of a meson.

\subsubsection{The bilocal fermionic constraints}

Having defined the necessary bilocal operators, 
we are able to write down a closed set of the bilocal 
fermionic constraints, which are obviously much involved than 
the previous case.
A closed set is formed by  ${\sf S,\, P}$ and the transverse 
components of vectors ${\sf V}^i$ and ${\sf A}^i$.
Indeed, the explicit form of the bilocal fermionic constraints 
are given by 
\begin{eqnarray}
 q^+ \pmatrix{{\sf S}(\p,\q)\cr {\sf P} (\p,\q ) }   
 &=&  \frac12 \left\{  \pmatrix{1\cr - i \gamma_5 }
       \left(\gamma^j q_\perp^j + m_0\right) \right\}_{\alpha\beta}
               {\sf M}_{\alpha\beta} (\p,\q )
    \nonumber \\
   &&-\, \frac{G_{\rm S}}{2}\int \frac{d^3 \k d^3 \bl}{(2\pi)^3}\, 
         {\sf M}_{\alpha\beta} (\p, \q -\k -\bl)\, 
         \biggl\{  \pmatrix{1\cr - i \gamma_5 }  \Big(  {\sf S}_{\rm R}
   + i \gamma_5 ~ {\sf P}_{\rm R}  \Big) \biggr\}_{\alpha\beta}(\k ,\bl)
       \nonumber \\
 && +\, \frac{G_{\rm V}}{2} \int \frac{d^3 \k d^3 \bl}{(2\pi)^3}
            \biggl[ 2\biggl\{ \Big(
              {\sf S} -i \gamma_5 \, {\sf P} \Big)
    \pmatrix{1\cr i \gamma_5 } \biggr\}_{\alpha\beta}(\p, \q-\k -\bl)
                 :{\sf M}_{\alpha\beta}(\k ,\bl):         \biggr.
\nonumber \\
 && \hskip0.4cm  \biggl.
          -\, {\sf M}_{\alpha\beta} (\p, \q -\k -\bl)\, 
   \biggl\{  \pmatrix{1\cr - i \gamma_5 } 
          \Big(  \gamma^j :{\sf V}^j_{\rm R}:
                + \gamma^j \gamma_5  \,  :{\sf A}^j_{\rm R}:  
          \Big)
                 \biggr\}_{\alpha\beta}
             (\k ,\bl)               \biggr],   \label{FC_full_1}   \\
 q^+\pmatrix{ {\sf V}^i (\p,\q ) \cr {\sf A}^i (\p,\q)} 
 &=&  -\frac12 \left\{  \pmatrix{ \gamma^i\cr -\gamma^i \gamma_5}
                      \left(\gamma^j q_\perp^j  +m_0\right)
                    \right\}_{\alpha\beta}
               {\sf M}_{\alpha\beta} (\p,\q )
\nonumber \\
   &&+\, \frac{G_{\rm S}}{2}\int \frac{d^3 \k d^3 \bl}{(2\pi)^3}\, 
         {\sf M}_{\alpha\beta} (\p, \q -\k -\bl)
         \biggl\{ \pmatrix{ \gamma^i\cr -\gamma^i \gamma_5}
          \Big( {\sf S}_{\rm R}
            + i \gamma_5 ~ {\sf P}_{\rm R}  
          \Big) \biggr\}_{\alpha\beta}(\k ,\bl)  \nonumber \\
 && +\, \frac{G_{\rm V}}{2} \int \frac{d^3 \k d^3 \bl}{(2\pi)^3}
            \biggl[ 2 \left\{
               \left( {\sf V}^i  + \gamma_5 \, {\sf A}^i \right)  
                \pmatrix{ 1 \cr \gamma_5} \right\}_{\alpha\beta}
                         (\p, \q -\k -\bl)
                 :{\sf M}_{\alpha\beta}(\k ,\bl):   \biggr.\nonumber \\
 && \hskip0.4cm  \biggl.
          +\, {\sf M}_{\alpha\beta} (\p, \q -\k -\bl)
       \biggl\{  \pmatrix{ \gamma^i\cr -\gamma^i \gamma_5} 
                 \left( \gamma^j  :{\sf V}^j_{\rm R}:  +
                        \gamma^j \gamma_5  \,  :{\sf A}^j_{\rm R}: 
                 \right)
       \biggr\}_{\alpha\beta} (\k ,\bl) \biggr]\, . \ \label{FC_full_2}
\end{eqnarray}
The plus components of the vectors ${\sf V}^+,\ {\sf A}^+$ do not 
contain the bad spinor and are expressed by ${\sf M}_{\alpha\beta}$,
while the minus components ${\sf V}^-,\ {\sf A}^-$ are expressed in 
terms of ${\sf S,\, P}$, ${\sf V}^i$ and ${\sf A}^i$.
We can solve the above equations following the same procedure 
as before, though all the calculations are tremendously cumbersome. 
It should be noticed that the leading order constraint for ${\sf S}$ 
again leads to the gap equation for the dynamical quark mass $M$, 
which is equivalent 
to Eq.~(\ref{gap}) because we took the normal order in the definition
of the vector interaction (\ref{vector_int}). 

\subsubsection{The LC Hamiltonian}


Substituting the solutions of the bilocal fermionic constraints 
into the canonical LC Hamiltonian, 
one finds again that the 
first nontrivial Hamiltonian starts from $n=2$ in the $1/N$ expansion 
(see, Eq.~(\ref{Ham_expand})). As a result, the Hamiltonian gets extra 
terms due to the vector interaction in addition to
 $h^{(2)}_{\rm S}$ given by (\ref{h2_wo_Vector}).  
After a lengthy calculation, one finally obtains (unimportant 
c-number is ignored)
\begin{eqnarray}
  h^{(2)}  \!\!     
  & =&\!\! h^{(2)}_{\rm S} \, 
       +\, \int  \frac{ [d \p ] [d \q ] [d \k ] [d  \bl]}{(2\pi)^3}\,
        \, \frac{\delta(  \p + \q - \k - \bl)}{ 4 \sqrt{p^+ q^+ k^+ l^+}} \NN
  && \hspace{1cm}\times  \sum_{s_1,s_2} \sum_{s'_1,s'_2} 
       \Big(\omega^{}_{\rm V} + \omega^{}_{\rm SV} 
         + \omega^{}_{\rm V^2} + \omega^{}_{\rm SV^2}\Big)
             B^\dag(\p, s_1 : \q , s_2 )     B(\k , s'_1 : \bl, s'_2 )\, .
\label{Hamiltonian_full}
\end{eqnarray}
The "kernel" $\omega$'s in the interaction terms are functions of 
momenta $\p,\q,\k,\bl$ and spins $s_1,s_2,s_1',s_2'$ and given by 
\begin{eqnarray}
&&\hspace*{-7mm}\omega^{}_{\rm V}\equiv g^{}_{\rm V}
  \biggl\{ \kappa^{}_{\rm V} \sum_{\mu=1,2}
                +  \sum_{\mu=+,-}  
  \biggr\}  
    \left[ 
    \Big\{ {\bar u} (\p,s_1) \gamma^\mu v(\q,s_2) \Big\}
    \Big\{ {\bar v} (\bl,s'_2) \gamma_\mu u(\k,s'_1) \Big\}
\right.    \nonumber   \\
 & &    \hskip6cm  \left.
+\, 
    \Big\{ {\bar u} (\p,s_1) \gamma^\mu \gamma_5 v(\q,s_2) \Big\}
    \Big\{ {\bar v} (\bl,s'_2) \gamma_\mu \gamma_5 u(\k,s'_1) \Big\}
   \right]    \, ,       \\
&&\hspace*{-7mm}\omega^{}_{\rm SV} \equiv i g^{}_{\rm S}g^{}_{\rm V}
 \kappa^{}_{\rm S}  \int \frac{d \bfb d \bfc }{(2\pi)^3} \,
             \frac{ \epsilon ( b^+) M}{b^+ c^+} \,
                \nonumber  \\
 & & \quad   \times \left[  \delta ( \p+\q+\bfb+\bfc )\, 
         \Big\{ {\bar u} (\p,s_1) \gamma^+ \gamma_5 v(\q,s_2) \Big\}
         \Big\{ {\bar v} (\bl,s'_2) i \gamma_5 u(\k,s'_1)\Big\}    
        \right.         \nonumber   \\
 & &    \hskip3.4cm 
\left. + \delta ( \k+ \bl-\bfb-\bfc ) \,  
          \Big\{ {\bar u}(\p,s_1) i \gamma_5 v(\q,s_2) \Big\}
         \Big\{ {\bar v} (\bl,s'_2) \gamma^+ \gamma_5 u(\k,s'_1) \Big\}  
           \right]      \, ,   \\
&& \hspace*{-7mm}\omega^{}_{\rm V^2} \equiv g_{\rm V}^2  
    \int \frac{d \bfb d \bfc }{(2\pi)^3} \,
         \frac{ \epsilon ( b^+) ( b^2_\perp +M^2 ) }{b^+ b^+ c^+} \, 
\delta ( \p+ \q -\bfb-\bfc )            \\
 & &   
\quad \times \left[    
         \Big \{ {\bar u} (\p,s_1) \gamma^+ v(\q,s_2) \Big\}
         \Big\{ {\bar v} (\bl,s'_2)  \gamma^+ u(\k,s'_1) \Big\} 
         +  
         \Big\{ {\bar u} (\p,s_1) \gamma^+ \gamma_5 v(\q,s_2) \Big\}
         \Big\{ {\bar v} (\bl,s'_2) \gamma^+ \gamma_5 u(\k,s'_1) \Big\}  
 \right]          \, , \NN
&&\hspace*{-7mm}\omega^{}_{\rm SV^2} \equiv g^{}_{\rm S}g_{\rm V}^2 
   \kappa^{}_{\rm S} 
   \int \frac{d \bfb d \bfc d \bfb' d \bfc' }{ (2\pi)^3} \,
                \frac{ \epsilon ( b^+) M}{b^+ c^+} \frac{
\epsilon ( b'^+) M}{b'^+ c'^+} \,  \\
 & &    \hskip2.0cm  \times   \delta ( \p+\q+\bfb+\bfc ) \, 
\delta (\k+\bl-\bfb' -\bfc' )       
       \Big\{ {\bar u} (\p,s_1) \gamma^+ \gamma_5 v(\q,s_2) \Big\}
       \Big\{ {\bar v} (\bl,s'_2) \gamma^+ \gamma_5 u(\k,s'_1) \Big\}   
 \, ,\nonumber
\end{eqnarray}
where we have introduced another type of divergent constant
\BQ\label{kappa_V}
 \kappa^{}_{\rm V}=\left\{ 1+g^{}_{\rm V}\int 
\frac{d^3\r}{(2\pi)^3} \frac{\epsilon(r^+)}{P^+-r^+}\right\}^{-1}\, .
\EQ
Notice that $\kappa^{}_{\rm V}$ is independent of 
the external momentum $P^+$ due to exactly the same reason as for 
$\kappa^{}_{\rm S}$ given by Eq.~(\ref{kappa_S}).

A few comments are in order about this LC Hamiltonian.
First of all, we have introduced the rescaled vector coupling constant 
$g^{}_{\rm V}=G_{\rm V}N$ which naturally yields the $1/N$ expansion 
of the LC Hamiltonian. 
Otherwise all the interaction terms become larger than $h_{\rm S}^{(2)}$
in the $1/N$ counting, which is not interesting. 
Next, it is clear that all the interaction terms are of the $B^\dag B$ 
type. This means that the Hamiltonian does not mix the Fock components 
with different number of particles. Namely, 
we can solve the Hamiltonian within a small Fock space
with one boson (or, equivalently, a $q\bar q$ pair). 

Lastly, it would be necessary to explain why we obtained terms 
like $\omega_{\rm SV}$ and $\omega_{\rm SV^2}$ which are of the 
higher order with respect to $G_{\rm S}$ and $G_{\rm V}$, 
which seems less trivial compared to the case without the vector 
interaction. The mathematical reasons are the following: 
First, the bilocal fermionic constraints (\ref{FC_full_1}) and 
(\ref{FC_full_2}) have contributions proportional to $G_{\rm S}$ or
$G_{\rm V}$. Thus, it is natural that the solutions to the fermionic 
constraints, in each order of the $1/N$ expansion, can depend upon 
$G_{\rm S}$ or $G_{\rm V}$. On the other hand, the LC Hamiltonian 
contains ${\sf V}^i_{\rm I}$, and this vector is again represented 
by other bilocal operators. Therefore, if we substitute the solutions 
(which depends on $G_{\rm S}$ and $G_{\rm V}$) to ${\sf V}^i_{\rm I}$ 
in the LC Hamiltonian,  there appear higher-order terms with respect to 
$G_{\rm S}$ or $G_{\rm V}$.

\section{Bound-state equations of light mesons}\label{Bound_State}
\setcounter{equation}{0}

\subsection{Interpolating fields and the lowest Fock state representation of mesons}

In the leading order of the $1/N$ expansion, mesonic states are 
expressed as "constituent" states with a dynamical quark-antiquark 
pair, as was discussed in Ref.~\cite{Itakura-Maedan_NJL}.
For the purpose of constructing a meson with appropriate statistics, 
the first thing to do is to determine its kinematical structure.
This can be done in such a way that the mesonic state has the same 
structure as that of its interpolating field. Namely, a part of 
the LC wavefunction which rather trivially depends on the statistics 
can be determined even without solving the dynamical bound-state 
equation. Then, after extracting this known part, we can define the 
spin-independent part of the LC wavefunctions.

As we already discussed, the interpolating fields are given by 
the "real" parts of the bilocal operators at the same space-time points
(see Eqs.~(\ref{Interpolating1}), (\ref{Interpolating2})).
We can find their explicit forms in terms of spinors and the 
bilocal boson operators, by using the solutions to the bilocal
fermionic constraints. Since (in this subsection) we are interested 
in the kinematical structures which is {\it independent of the 
dynamics}, we are allowed to use the solutions to the fermionic 
constraint without the vector interaction\footnote{In fact, we can 
even use the free theory (a free fermion with dynamical mass $M$) 
to determine the kinematical structure. 
However, we do not perform this because we already have the solutions 
to the case with ${\cal L}_{\rm S}\neq 0$.}.
To the leading order, one finds for the scalar and pseudo-scalar fields
\begin{eqnarray}
\pmatrix{\bar\Psi (\x)  \Psi (\x)\cr
 \bar\Psi (\x)  i \gamma_5 \Psi (\x)  }
  & = & \pmatrix{ {\sf S}_{\rm R}( \x, \x)\cr
{\sf P}_{\rm R}( \x, \x)   }  \nonumber \\
 &=&  ~\sqrt{N} \int \frac{[d  \PP]}{(2\pi)^3}  
\left[\ {\rm e}^{ - i \PP  \x} ~ \int_0^{P^+} d k^+ \int d^2 k_\perp
           \frac{\kappa^{}_{\rm S}}{2 \sqrt{  k^+ (P-k)^+ } }\right.
 \label{inter1}\\
&& \left.\times \sum_{s_1, s_2}
  \Big\{ {\bar u} (\k,s_1) \pmatrix{1\cr i\gamma_5}v(\PP - \k,s_2)\Big\}
   B^{\dag} ( \k, s_1 : \PP  - \k, s_2 )\ + {\rm h.c.}\ 
                \right]        + {\cal O}( N^0 )\, ,\nonumber
\end{eqnarray}
and for the vector 
\begin{eqnarray}
\bar\Psi(\x)  \gamma^\mu  \Psi (\x) 
  & = &  
{\sf V}^{\mu}_{\rm R}( \x, \x) 
\nonumber \\
 &=&  ~\sqrt{N} \int\frac{[d  \PP ]}{(2\pi)^3} 
 \left[ \, {\rm e}^{ - i \PP  \x} ~ \int_0^{P^+} d k^+ \int d^2 k_\perp
           \frac{1}{ 2\sqrt{ k^+ (P-k)^+ } }\right. \\
 & &    \left.  \times   \sum_{s_1, s_2}
              \Big\{ {\bar u} ( \k,s_1)  
\gamma^\mu 
v(\PP  - \k, s_2) \Big\} B^{\dag} ( \k, s_1 : \PP  - \k, s_2 )
        + {\rm h.c.}        \right]  + {\cal O}( N^0 )\, .\nonumber
\end{eqnarray}
The four interpolating fields of the vector 
are obtained separately for the LC components $\mu=\pm$ and 
the transverse components $\mu=1,2$, but it turns out that they can be 
combined to give the reasonable representation as above.  
The mismatch in the expressions above between $({\sf S,P})$ and 
${\sf V}$ (namely, a factor $\kappa^{}_{\rm S}$ in Eq.~(\ref{inter1})) 
is due to our use of the solutions with ${\cal L}_{\rm S}\neq 0$ and 
${\cal L}_{\rm V}= 0$, and would be absent if one used the free field
solutions.  Needless to say, this difference is irrelevant for the 
present purpose of determining the kinematical structure.

Since the interpolating fields are all defined at a single spatial point, 
they can be interpreted as a kind of point-like states without spatial 
structure. However, what we are interested in is the spatial structure 
of mesons, and it is indeed the (spin-independent part of the) LC 
wavefunction which is responsible for this spatial structure.
(In other words, for the interpolating fields without spatial structure, 
the LC wavefunctions are simply constant in momentum space.) 
First of all, the mesonic states having total momentum $\PP$ 
can be represented as follows ($x=k^+/P^+$):
\BQ
 \ket{\, {\rm meson}\, ; \PP \, } =  
P^+  \int_{0}^{1} {dx}\int d^2 k_\perp ~ \sum_{s_1,s_2}
\Phi_{\rm meson} ( x, k_{\perp}, s_1, s_2 )  
B^\dag(\k , s_1 : \PP -\k, s_2 )  \ket{0}\, ,\label{meson_generic}
\EQ
where $\Phi_{\rm meson} ( x, \k_{\perp}, s_1, s_2 )$ is the LC 
wavefunction of the $q\bar q$ Fock state, and the vacuum is 
defined by the bilocal boson operator 
\BQ
B(\p,s_1:\q,s_2)\ket{0}=0\, .
\EQ
Notice that this vacuum is equivalent to the one defined 
by the quark antiquark operators
to the leading order of the $1/N$ expansion.
From the interpolating fields shown above, we can determine the
kinematical structure (or, spin-dependent parts) of the LC 
wavefunctions:
\begin{eqnarray}
     \Phi_\pi ( x, k_{\perp}, s_1, s_2 )  & \equiv &
           \frac{1}{ 2 \sqrt{  x(1-x)} } \,  \phi_\pi (x, k_{\perp }  ) \,
     \Big\{ {\bar u} ( \k,s_1) \, i \gamma_5 \, v( {\PP}-\k, s_2) \Big\},
 \label{Pi_state} \\
    \Phi_\sigma ( x, k_{\perp}, s_1, s_2 )  & \equiv &
           \frac{1}{ 2 \sqrt{  x(1-x)} } \,  \phi_\sigma (x, k_{\perp }  ) \,
             \Big\{ {\bar u} ( \k,s_1)  \, v( {\PP}-\k, s_2) \Big\},
       \label{Sigma_state}  \\
    \Phi_\rho^\lambda ( x, k_{ \perp},  s_1, s_2 )
    & \equiv &
          \frac{1}{ 2 \sqrt{  x(1-x) } } \,  \phi_\rho (x, k_{ \perp }  ) \,
            \epsilon_\mu ( \lambda, P )   
          \Big\{ {\bar u} ( \k,s_1) \, \gamma^\mu \,
                   v(  {\PP}-\k, s_2) \Big\}\, ,
 \label{Vec_state}
\end{eqnarray}
where we have introduced the spin-independent LC 
wavefunction\footnote{The spin-independent LC wavefunction in the present
paper is different from that of the previous papers 
\cite{Itakura-Maedan_NJL,Itakura-Maedan_review} by a factor $2x(1-x)$.}
$\phi(x,k_\perp)$ for each meson.
These expressions are valid when $P_\perp =0$ . The corresponding formulae
for $P_\perp\neq 0$ are shown in Appendix~\ref{LCWF_NonzeroP}.
For the vector meson, we have already introduced 3 physical states 
$\ket{\rho,\lambda}$ with helicity $\lambda=\pm1,0$ (see Eq.~(\ref{Vec_state}))
which should be
distinguished from 4 states $\ket{\rho^\mu}$ naively constructed 
from the interpolating field without the polarization vector 
$\epsilon_\mu(\lambda,P)$. 
Below, we will explain the necessity of using $\ket{\rho,\lambda}$
instead of $\ket{\rho^\mu}$.

One can directly verify the orthogonality among these mesonic states.
For example, let us check the orthogonality between the scalar $\ket{\sigma}$
and the vector $\ket{\rho^\mu}$ when the transverse momentum
of the meson $P_\perp$ is zero:
\begin{eqnarray}
&&\hspace*{-2cm}
\bra{\, \sigma ; P^+, P_\perp =0 \,} \, \rho^\mu ; P'^+, P'_\perp =0\, \rangle
\NN
  &=& \delta ( \PP - \PP' )\vert_{P_\perp=0}\,  P^+
\int_0^1 \frac{dx}{4 x (1-x) } \int_{-\infty}^{\infty} d^2 k_\perp
         { \phi}^*_\sigma ( x, k_\perp )  { \phi}_\rho ( x, k_\perp )
            \nonumber \\
  & &  \hskip1cm \times   \sum_{s_1,s_2} ~ \Big\{ {\bar u} ( \k,s_1) \,
\gamma^\mu \, v( \PP -\k, s_2) \Big\}
     \Big\{ {\bar v} ( \PP -\k,s_2)  \, u( \k , s_1) \Big\} 
\Big\vert_{P_\perp=0}\,  ,
\end{eqnarray}
where the quantity on the last line can be evaluated as
\begin{eqnarray}
&&\hspace*{-2cm}\sum_{s_1,s_2} ~ \Big\{ {\bar u}(\k,s_1)\, 
       \gamma^\mu \, v( \PP-\k, s_2) \Big\}
  \Big\{ {\bar v}(\PP -\k,s_2)\, u(\k , s_1)\Big\}\Big\vert_{P_\perp=0}\NN
  &=&    \left\{  
\begin{array}{ll}
         -8M k^i_\perp  \, ,      &  \; \mbox{  $ \mu = i = 1,2 $  } \\
          4MP^+ (1-2x)   \, ,      &  \; \mbox{  $ \mu = + $ }     \\
       -\frac{4M}{2P^+} \frac{(1-2x)}{x(1-x)} \,  ( k^2_\perp + M^2 ) 
              \, ,               &  \;  \mbox{ $ \mu = - $  } .
\end{array}
     \right.
\end{eqnarray}
Assuming that the spin-independent wavefunctions are functions 
of $x(1-x) $ and $ k_\perp^2 $ (which 
can be checked  a posteriori, but is natural due to the charge symmetry), 
one concludes that the bracket $\bra{\sigma}\rho^\mu\rangle$ 
above is zero for all $\mu$.
The orthogonality for other combinations of the states are 
verified too in a similar way.

It should be noticed that the four vector states $\ket{\rho^\mu}$ 
are not independent of each other (thus not necessarily orthogonal 
to each other),
since a real massive vector meson should have only three physical modes, 
i.e., two transverse and one longitudinal modes. 
This can be seen by the fact that the 
spinor part of the state (\ref{Vec_state}) 
satisfies the following identity:
\begin{eqnarray}
   & & (E_k +E_{(P-k)} ) \left[ {\bar u}( \k, s_1 ) \gamma^+  \, 
                                 v(\PP-\k, s_2 ) \right]
     +P^+  \left[ {\bar u}( \k, s_1 ) \gamma^-  \,
                  v( \PP -\k, s_2 ) \right] \NN
   && - P_\perp ^i \left[ {\bar u}( \k, s_1 ) \gamma^i  \, 
                        v( \PP -\k,s_2 ) \right] =0     \nonumber 
\end{eqnarray}
where $E_k$ and $E_{P-k}$ are the LC energies of the quark and the 
antiquark
\begin{eqnarray}
  E_{k} \equiv \frac{k_\perp ^2 +M^2}{2 k^+},\qquad 
  E_{(P-k)} \equiv \frac{(P-k)_\perp ^2 +M^2}{2 (P-k)^+}\, .
\end{eqnarray}
The projection onto 
three physical modes can be easily done by the 
circular polarization vector $\epsilon^\mu ( \lambda, P )$. 
As we already introduced in Eq.~(\ref{Vec_state}), three independent 
states with helicity $\lambda=\pm 1, 0$ are given by 
\begin{equation}
   \ket{\, \rho, \lambda\, ; \PP\, }\equiv 
    \epsilon_\mu (\lambda, P)  \ket{\, \rho^\mu\, ; \PP \, }.
  \label{Physical_vector_state}
\end{equation}
These three states are orthogonal:
$\bra{\, \rho, \lambda\,  }\, \rho, \lambda'\, \rangle =0$
for $\lambda\neq \lambda'$.
In the frame where the transverse momentum of the vector meson is zero, 
    $ \PP  =( P^+, P_\perp ) = (   P^+,  0_\perp )$,
the polarization vectors are given by
{\bf (}$\epsilon^\mu=(\epsilon^+,\epsilon^-,\epsilon^1,\epsilon^2)$
{\bf )} \cite{Form_Factor}
\begin{eqnarray}
    \epsilon^\mu ( \lambda=\pm1, P )  &=&
              (0, 0, \mp \frac{1}{ \sqrt{2} } , -\frac{i}{\sqrt{2} } )\, , \\
   \epsilon^\mu ( \lambda=0, P )  &=&
      (\frac{P^+}{ m^{}_{\rm V}}, -\frac{m^{}_{\rm V}}{ 2 P^+ }, 0, 0 )\, ,
  \label{polarization}
\end{eqnarray}
satisfying $ \epsilon_\mu (\lambda, P ) \epsilon^\mu (\lambda', P )^*
=-\delta_{\lambda \lambda'},~
                    P_\mu \epsilon^\mu (\lambda, P ) =0$.
Therefore, the explicit relations between $\ket{\rho^\mu}$ and the physical
states $\ket{\rho,\lambda}$ are
\begin{eqnarray}
   & &   \ket{\,\rho,\lambda=\pm 1\,}
           = \pm \frac{1}{\sqrt{2}} \, \vert \, \rho^1\,\rangle
           + \frac{i}{\sqrt{2}} \, \vert \, \rho^2\,\rangle\, ,
         \label{Trans_state} \\
  & &  \ket{\,\rho,\lambda=0\,}
           = \frac{-m^{}_{\rm V}}{ 2 P^+ } \, \vert \, \rho^+\,\rangle
           + \frac{P^+}{ m^{}_{\rm V} } \, \vert \, \rho^-\, \rangle \, .
  \label{Long_state}
\end{eqnarray}
Notice that these relations are valid only in the frame 
$ \PP = (   P^+,  0_\perp )$,
and does not hold any longer if one moves to other reference frame.

In summary, the orthogonality holds among the five mesonic states:
$\ket{\sigma},\, \ket{\pi}$ and $\ket{\rho,\lambda}$
with $\lambda=\pm1,0$. Therefore, together with the normalization 
of each state, we obtain 
\BQ
\langle \alpha, \PP \ket{\alpha', \PP'}= 
16\pi^3 P^+  \, \delta_{\alpha,\alpha'}\, \delta (  {\PP}- {\PP'} ),
\EQ
where $\alpha$ ($\alpha'$) distinguishes the species of meson 
$\alpha=\sigma,\pi,\, (\rho,\lambda)$.
This also leads to the normalization condition for 
the LC wavefunctions:
\begin{eqnarray}
 \frac{1}{ 16 \pi^3 } \int_{0}^{1} dx  \,
         \int_{-\infty}^{\infty} d^2 k_{\perp } ~\sum_{s_1,s_2}
          \vert  \Phi_\alpha  ( x, k_{ \perp},  s_1, s_2 )  \vert^2 =1.
\label{normalization}
\end{eqnarray}

\subsection{Bound-state equations}

Before going into details, let us briefly discuss a generic 
$q\bar q$ state in order to clarify the procedure we perform.
In general, the LF energy of the two body state 
may be schematically written as 
\BQ
P^-_{q\bar q} = \frac{\kk^2+M^2}{2k^+} + 
\frac{(P_\perp-\kk)^2 + M^2}{2(P^+-k^+)} + V(k,P),
\EQ
where the first two terms are the "kinetic" energies of the quark and 
the antiquark, and $V$ is the potential which allows for a bound state.
This form of the energy leads to the following bound-state equation:
\BQ
\left\{m_{\rm meson}^2 - \frac{\kk^2+M^2}{x(1-x)}\right\}\phi(x,\kk)=
\int_0^1dy \int d^2p_\perp \, V(x,\kk;y,p_\perp)\, \phi(y,p_\perp)\, ,
\label{LFBSE_general}
\EQ
where we have chosen 
$\Kmbf{P}= ( P^+, P_\perp^i )  = ( P^+,\, 0_\perp ) $
 for simplicity, and redefined $V$ with some factors included. 

For the scalar and pseudo-scalar mesons, the potential $V$ has 
a simple structure, and we can evaluate it without any further approximation.
On the other hand, for the vector mesons, $V$ has nontrivial dependencies
upon $g^{}_{\rm V}$. 
For the present purpose of seeing the leading effects of 
the vector interaction (\ref{vector_int}) on the vector channel, 
 it is actually enough to derive the potential $V$ 
up to the leading order of the vector interaction $g^{}_{\rm V}$. 
We will see that in this leading order 
the potential term $V(x,\kk;y,p_\perp)$ is separable with respect to 
the internal $(y,p_\perp)$ and external $(x,\kk)$ variables, 
and actually depends only on $y$ and $p_\perp$. 

In the following, we are going to derive a bound-state equation 
for each meson in the frame $\Kmbf{P}  = ( P^+,\, 0_\perp )$, 
which is simply deduced from the light-front eigen-value equation:
\begin{equation}
  P^- \ket{\,{\rm meson}\,; P^+,P_\perp=0 }~
     =\frac{m_{\rm meson}^2}{2P^+} \ket{\,{\rm meson}\,; P^+,P_\perp=0 }\, .
\label{LFBSE}
\end{equation}
To the first non-trivial order in the $1/N$ expansion, this equation 
can be restricted to the $q\bar q$ subspace of the Fock space, leading to
the bound-state equation for mesons.

\subsubsection{Scalar and pseudo-scalar mesons}
It is straightforward to derive the bound-state equations for scalar
and pseudo-scalar mesons. In Eq.~(\ref{LFBSE}), we use the LC 
Hamiltonian $P^-=h^{(2)}_{\rm S}$ defined in Eq.~(\ref{h2_wo_Vector}) 
(recall our strategy  for scalar and pseudo-scalar mesons, see 
Sect.~\ref{NJL})
and the scalar and pseudo-scalar states given by
Eqs.~(\ref{Sigma_state}), (\ref{Pi_state}). Then, one obtains
\begin{eqnarray}
&&\hspace*{-5mm}
\left\{m_\sigma^2 - \frac{k_\perp^2+M^2}{x(1-x)}\right\} \phi_\sigma(x,k_\perp)
= -  \frac{g^{}_{\rm S}\kappa^{}_{\rm S}}{(2\pi)^3}
    \int_0^1 dy  \int d^2 p_\perp
      \frac{p_\perp^2 + (1-2y)^2 M^2}{y^2(1-y)^2}
    \phi_\sigma (y , p_\perp) \, , \label{LFBSE_sigma}\\
&&\hspace*{-5mm}
\left\{ m_\pi^2 -  \frac{k_\perp^2+M^2}{x(1-x)}\right\} \phi_\pi (x, k_\perp)
 =-  \frac{g^{}_{\rm S}\kappa^{}_{\rm S}}{(2\pi)^3}
    \int_0^1 dy  \int d^2 p_\perp
      \frac{p_\perp^2 + M^2}{y^2(1-y)^2}
    \phi_\pi (y , p_\perp) \, . \label{LFBSE_pion}
\end{eqnarray}
An important property to be noticed is the fact that, in both cases,
 the potential terms (the right-hand sides) are {\it independent} of 
the external variables $x,\ k_\perp$. This property is due to the 
simple structure of the {\it contact} interaction ${\cal L}_{\rm S}$. 
The same thing will be seen in the case of the vector meson. 
These results are equivalent to the previous ones obtained in 
Ref.~\cite{Itakura-Maedan_NJL}.
Formal solutions to these equations are easily obtained (due to the 
simple structure of the potential terms), but we postpone to 
discuss them until the next section, where we present the LC 
wavefunctions and masses of the mesons after careful evaluation 
of divergences.

\subsubsection{Vector mesons}
For a vector meson, the first nontrivial contribution 
(in the $1/N$ expansion) of the LC Hamiltonian is given by 
$P^-=h^{(2)}$ in Eq.~(\ref{Hamiltonian_full}), and 
the transverse and longitudinal states are given by 
Eqs.~(\ref{Trans_state}) and (\ref{Long_state}), respectively. 
Although the mass and the {\it spin-independent} LC wavefunction 
must be, by definition, the same for both the transverse and 
longitudinal polarizations,
we distinguish the bound-state equations for these two modes 
since they appear differently until we perform careful regularization 
of the divergences.
Therefore, in the following, we use notations like 
$V_{\rm T/L},\ \phi_{\rm T/L}(x,k_\perp),$ and 
$m_{\rm T/L}$ for corresponding 
transverse and longitudinal quantities 
(though eventually they are proven to be equivalent to each other).

For a transversely polarized vector meson, 
a lengthy calculation yields the following potential $V_{\rm T}$
(in the sense of Eq.~(\ref{LFBSE_general}))
\begin{eqnarray}
V_{\rm T}
 =  -\, \, \frac{ g^{}_{\rm V}  }{(2\pi)^3}  
\left[1  + \frac{g^{}_{\rm V}}{(2\pi)^3} \int_{-\infty}^{\infty} dq^+
 \int d^2q_\perp \frac{ \epsilon(q^+) }{ P^+ - q^+ } \right]^{-1}
 \frac{ p_\perp^2 + M^2 - 2y(1-y) p_\perp^2 }{y^2(1-y)^2}\, .
\label{VT}
\end{eqnarray} 
This potential comes from $\omega^{}_{\rm V}$ 
in the LC Hamiltonian (\ref{Hamiltonian_full}).
The other terms do not contribute to the bound-state equation 
because of the orthogonality among
the bispinors.
Notice that the potential  is already 
independent of the external variables $x, \kk$. 
Taking the leading contribution of $g^{}_{\rm V}$, 
one arrives at an equation for the LC wavefunction 
$\phi^{}_{\rm T}(x,k_\perp)=\phi_{\lambda=\pm 1}(x,k_\perp)$:
\begin{eqnarray}
\left\{  m_{\rm T}^2  - \frac{ k_\perp^2 + M^2}{x(1-x)} \right\} 
 \phi^{}_{\rm T}(x,k_\perp)  
 =  -\, \frac{ g^{}_{\rm V}  }{(2\pi)^3}  \int_0^1 dy\int d^2 p_\perp
 \frac{ p_\perp^2 + M^2 - 2y(1-y) p_\perp^2 }{y^2(1-y)^2}\,
 \phi^{}_{\rm T}(y,p_\perp).
 \label{EQ:TRANS}
\end{eqnarray} 
Since we pick up the leading contribution in Eq.~(\ref{VT}), 
the divergent factor in the square bracket is now replaced by 
1. This factor is the explicit expression of $\kappa^{}_{\rm V}$ 
which appeared in the Hamiltonian (cf: see Eq.~(\ref{kappa_V})), and 
if we keep this factor, the structure of the bound-state equation 
looks similar to those of the scalar and pseudo-scalar mesons 
(their potential terms are proportional to $g^{}_{\rm S}\kappa^{}_{\rm S}$). 
However, as we will see below, 
the corresponding equation for the longitudinal mode does not
have the same simple structure, and we keep the leading order 
effect (namely, $\kappa^{}_{\rm V}=1$) for simplicity.

The longitudinally polarized state is much more involved. 
One of the main complexity comes from the fact that 
${\sf V}^+$ and ${\sf V}^-$ are treated (and in fact do behave) 
differently in the light-front quantization, while the physical 
longitudinal mode is a linear combination between these two. 
This situation requires a longer and attentive, but still 
straightforward, calculation which eventually leads to the following
potential $V_{\rm L}$ (for the details of the 
derivation, see Appendix \ref{BSELONG}):
\begin{eqnarray}
V_{\rm L} 
  &\! \! =&\!\! - \frac{ g^{}_{\rm V} }{(2\pi)^3  }
          \left\{ 1 - 2 \frac{ g^{}_{\rm V} }{(2\pi)^3  } 
          \int_0^1 dz \int d^2 q_\perp \right\}^{-1}      
         \left\{m_{\rm L}^2 - \frac{k_\perp^2 + M^2}{x(1-x)}\right\} 
         \left\{m_{\rm L}^2 + \frac{k_\perp^2 + M^2}{x(1-x)}\right\}^{-1} 
      \NN
 &  \times &\hspace*{-3mm}
    \left[\, 2 \, + \, \frac{4 \, (k_\perp^2+M^2)}
                            {m_{\rm L}^2 \, x(1-x) - (k_\perp^2+M^2 ) }
    \left\{1 - \frac{ g^{}_{\rm V} }{(2\pi)^3} 
          \int_0^1\! dz\! \int\! d^2 q_\perp \right\} \right]
    \left\{m_{\rm L}^2 \, + \frac{p_\perp^2 + M^2 }{y(1-y)} \, 
    \right\} . 
         \label{EQ:STLONG}
\end{eqnarray}
It should be noticed that the interaction term $\omega^{}_{\rm V^2}$ 
in the LC Hamiltonian (\ref{Hamiltonian_full}) as well as  
$\omega^{}_{\rm V}$ contributes to the longitudinal potential 
while it does not contribute to the transverse potential.
As we mentioned above, the $g^{}_{\rm V}$ dependence is not the same 
as the transverse case, and is much more non-trivial. 
However, we obtain a simpler result by taking the "leading term"
with respect to $g^{}_{\rm V}$ in this potential (more precisely, 
in the eigen-value equation). Here we give a quick, but less systematic, 
derivation of the final result. (A more detailed derivation is
 explained in Appendix \ref{BSELONG}.)
This can be done as follows:
First of all, we ignore the $g^{}_{\rm V}$ dependent term in the second line 
of Eq.~(\ref{EQ:STLONG}) that gives the higher order in 
$g^{}_{\rm V}$ and thus 
can be ignored anyway. We immediately find
$$
\left\{  m_{\rm L}^2  - \frac{ k_\perp^2 + M^2 }{x(1-x)} \right\} 
\left\{  m_{\rm L}^2  + \frac{ k_\perp^2 + M^2 }{x(1-x)} \right\}^{-1} 
\left[ \, 2 \,  +  \, \frac{ 4 \,( k_\perp^2+M^2 )}
                           {m_{\rm L}^2  \, x(1-x) -(k_\perp^2+M^2 ) }
\right]=2\, .
$$
Then we integrate the resulting bound-state equation over $x$ and $\kk$, 
obtaining the following:
$$
\int dx\, d^2\kk  \left\{  m_{\rm L}^2  - \frac{ k_\perp^2 + M^2 }{x(1-x)} 
\right\} 
 \phi^{}_{\rm L}(x,k_\perp) = -4 \frac{g^{}_{\rm V}}{(2\pi)^3}
\left(\int dx\, d^2\kk\right)\int dy\, d^2 p_\perp 
  \frac{p_\perp^2+M^2}{y(1-y)}\, \phi^{}_{\rm L}(y,p_\perp).
$$
Here we have kept the $g^{}_{\rm V}$ dependent term in the curly bracket 
in the first line of Eq.~(\ref{EQ:STLONG}).
We can modify the bound-state equation by using this relation in the 
right-hand side of it. Taking the leading term with respect 
to $g^{}_{\rm V}$, we finally obtain
the following simpler equation
for the longitudinal mode $\phi^{}_{\rm L}(x,k_\perp)=
\phi_{\lambda=0}(x,k_\perp)$:
\begin{eqnarray}
 \left\{  m_{\rm L}^2  - \frac{ k_\perp^2 + M^2 }{x(1-x)} \right\} 
 \phi^{}_{\rm L}(x,k_\perp) 
 =  - \, \frac{ g^{}_{\rm V} }{(2\pi)^3}
  \int_0^1 dy\int d^2 p_\perp
 \frac{ 4(p_\perp^2 + M^2) }{y(1-y) }\, \phi^{}_{\rm L}(y,p_\perp)\, .
 \label{EQ:LONG}
\end{eqnarray} 
It is evident that the right-hand side is again 
independent of the variables $x,\kk$. The derivation of 
Eq.~(\ref{EQ:LONG}) looks tricky, but we can obtain the same result by 
carefully taking the leading order effect with respect to $g^{}_{\rm V}$
in the eigen-value equation (see Appendix~\ref{BSELONG}).

Now we obtained the bound-state equations for the transverse and 
longitudinal modes (i.e., Eqs.~(\ref{EQ:TRANS}) and (\ref{EQ:LONG}) ).
However, at first glance, these two look different from each other
and thus seem to give different masses for the transverse and 
longitudinal vector mesons.
This is of course physically unacceptable, and as we will verify soon, 
these equations are in fact identical and give the same mass 
$m^{}_{\rm T}=m^{}_{\rm L}$. This equivalence will be achieved 
after one specifies cutoff scheme.
It is not hard to identify the origin of this (fake) discrepancy with 
the lack of Lorentz covariance in the light-front formalism. 
Namely, the main source is the fact that the 
LC components ${\sf V}^\pm$ and the transverse components ${\sf V}^i$ 
seem to behave differently in the light-front quantization.

\section{LC wavefunctions and masses of light mesons}\label{LCWF_mass}
\setcounter{equation}{0}

\subsection{General arguments: Eigenvalue equation and the cutoff scheme}
In the previous section, we have found that all the bound-state 
equations have the potential terms which do not depend upon the 
external variables $x,\kk$.\footnote{This will be due to the contact 
interaction in the NJL model, but actually, within the present formalism,
it appears in a highly nontrivial way, especially for the vector mesons.}
This property leads to a significant consequence that the spin-independent
LC wavefunction $\phi(x,\kk)$ should have the same form as 
a function of $x,\kk$. Let us see this in general arguments. 
Consider Eq.~(\ref{LFBSE_general}) with such a "separable" potential
(the subscript $i$ stands for $\sigma,\pi,$ T, L):
$$
\left\{m_i^2 - \frac{\kk^2+M^2}{x(1-x)}\right\}\phi_i(x,\kk)=
\int_0^1dy \int d^2p_\perp \, V_i(y,p_\perp)\, \phi_i(y,p_\perp)\, .
$$
Since the right-hand side is just a constant, one immediately finds  
that the spin-independent LC wavefunctions should have the common 
functional dependence upon $x$ and $\kk$ as we announced in the introduction:
\BQ
\phi_i(x,\kk)=\frac{C_i}{m_i^2-\frac{\kk^2+M^2}{x(1-x)}}\, .
\label{Solution_general}
\EQ
The constant factor $C_i$ simply represents the right-hand side of 
the bound-state equation,
\BQ
C_i=\int_0^1dy \int d^2p_\perp \, V_i(y,p_\perp)\, \phi_i(y,p_\perp)\, ,
\EQ
and is determined from the normalization condition of the wavefunction 
(\ref{normalization}). Note that our LC wavefunction 
(\ref{Solution_general}) is the simplest wavefunction from the general 
consideration of the relativistic bound-state equation \cite{BL-review,Heinzl};
All the LC wavefunctions must have the similar denominator as in 
Eq.~(\ref{Solution_general}), while
the numerator could be a function of $x$ or $\kk$ in general.
The spin-independent LC wavefunction always has a peak at $x=1/2$, 
but its shape 
(width of the peak) will change depending on the value of mass $m_i$. 
Namely, the peak becomes sharp as $m_i$ becomes large.

Substituting Eq.~(\ref{Solution_general}) into the above definition of 
$C_i$, one obtains an equation for $m_i$ (the eigen-value equation):
\BQ
1=\int_0^1dy \int d^2p_\perp \, 
\frac{V_i(y,p_\perp)}{m_i^2-\frac{p_\perp^2+M^2}{y(1-y)}}\, \, .
\label{Eigen_eq_general}
\EQ
Though all the potential terms $V_i$ ($i=\sigma,\, \pi,$ T, L) 
are always negative,  this equation makes sense because 
the denominator can be either positive or negative.  

Since the integral (\ref{Eigen_eq_general}) is divergent in general,
we need to regularize it. 
We evaluate the integral
by introducing the "extended parity invariant cutoff" 
\cite{Itakura-Maedan_NJL,PLB}
which is actually equivalent to the Lepage-Brodsky 
cutoff \cite{Brodsky-Lepage}: 
\begin{equation}
    \frac{  p_\perp^2 + M^2  }{ y(1-y)  } < 2 \, \Lambda^2\, .
    \label{LB}
\end{equation}
Indeed, this is a natural extension of the parity invariant cutoff 
in the {\it two body} sector, $K^+ K^- <\Lambda^2$ 
where $K^+$ and $K^-$ are the sum of 
(on-shell) quark and antiquark longitudinal momenta and energies 
[$K^+=p^++(P^+-p^+)=P^+,\ 
  K^-=(p_\perp^2+M^2)/2p^+ +(p_\perp^2+M^2)/2(P^+-p^+)$ ].
This cutoff obviously preserves transverse rotation and parity 
symmetry separately, but in fact it does work better. First, it also 
respects the usual three dimensional space rotation \cite{Bentz}. 
Thus one can relate the above cutoff $\Lambda$ to the 3-momentum 
cutoff $\sum_{i=1,2,3}(p^i)^2 < 
\Lambda^2_{\rm 3M}$ through $ 2(\Lambda^2_{\rm 3M}+M^2) = \Lambda^2$.
Next, the cutoff $K^+K^-<\Lambda^2$ is invariant under 
the boost transformation $K^\pm\to \, {\rm e}^{\pm\beta}K^\pm$, 
which is necessary for the relativistic formulation
\footnote{If one imposed the same condition $|K^\pm|<\Lambda$ as 
the one for the gap equation, it would violate the boost invariance,
though yielding the same Lepage-Brodsky cutoff. 
However, as we mentioned before, there is no problem in putting 
$|p^\pm|<\Lambda$ in the gap equation because the momentum is not the 
external momentum but the internal one to be integrated out.}.
Therefore, we are going to evaluate the following regularized 
eigenvalue equation: 
\begin{eqnarray}
1=\int_{y_-}^{y_+} dy  \int_0^{2 \Lambda^2 y (1-y)-M^2 } 
 \pi  \, d (p_\perp^2) \, 
 \frac{V_i(y,p_\perp)}{m_i^2-\frac{p_\perp^2+M^2}{y(1-y)}}\, \, .
\label{Eigen_eq_general_regulated}
\end{eqnarray}
where
\begin{equation}
  y_\pm = \frac{1 \pm \beta}{2},  \hskip1cm  \beta \equiv 
\sqrt{ 1- \frac{2M^2}{\Lambda^2} }.\label{y_beta}
\end{equation}
Below, we will see that this cutoff derives reasonable results for the
masses of mesons.

\subsection{Scalar and pseudo-scalar mesons}

We evaluate the eigenvalue equation 
(\ref{Eigen_eq_general_regulated})
for the scalar and pseudo-scalar potentials:
$$
V_\sigma= -\frac{g^{}_{\rm S}\kappa^{}_{\rm S}}{(2\pi)^3}
\frac{p_\perp^2+(1-2y)^2M^2}{y^2(1-y)^2}\, , \qquad 
V_\pi=-\frac{g^{}_{\rm S}\kappa^{}_{\rm S}}{(2\pi)^3}
\frac{p_\perp^2+M^2}{y^2(1-y)^2}\, .
$$
When we treat Eq.~(\ref{Eigen_eq_general_regulated}), we 
encounter two types of divergence in the longitudinal integrals. 
One is of the type $\int_0^1 dy/y$ which appears as the overall $y$ 
integral in Eq.~(\ref{Eigen_eq_general_regulated}),
and the other is $\int_0^\infty dy/y$ from the factor 
$\kappa^{}_{\rm S}$ in the potentials. The first type of the 
divergence is regulated by using the extended parity invariant cutoff 
as indicated in Eq.~(\ref{Eigen_eq_general_regulated}). 
On the other hand, we already know how to regulate the second 
type of the divergence, since it appeared in the gap equation
(\ref{gap_integral}). However, in fact, we do not have to know 
the explicit expression of the regulated integral (regulated 
by the parity invariant cutoff) for the present purpose. 
We rather utilize the gap equation (\ref{gap_integral}) 
to replace the divergent integral by a finite quantity. 
Indeed, the divergent factor $\kappa^{}_{\rm S}$ is modified 
so that it contains only the first type of the divergence:
\BQA
\kappa^{-1}_{\rm S}&=& 1+ \frac{g^{}_{\rm S}}{(2\pi)^3}\int {d^2 q_\perp}
\left[\int_0^\infty \frac{dz}{1-z}- \int_{-\infty}^0 \frac{dz}{1-z}
      \right] \NN
&=&1+ \frac{g^{}_{\rm S}}{(2\pi)^3}\int d^2q_\perp \left[
\left(\int_0^1 \frac{dz}{z}
-\int_0^\infty \frac{dz}{z}\right) 
-\left(-\int_0^1 \frac{dz}{z} + \int_0^\infty \frac{dz}{z}
\right)\right]\NN
&=&\frac{m_0}{M}+  \frac{ 2g_{\rm S} }{ (2\pi)^3} \int_{0}^{1}
\frac{d z}{z} \int d^2 q_\perp,
\EQA
where we have used the gap equation (\ref{gap_integral}) in the last line.
For the remaining longitudinal integral, 
 we can use the same cutoff as indicated in 
Eq.~(\ref{Eigen_eq_general_regulated}). 
For example, the equation for the pion is evaluated as follows:
\begin{eqnarray}
0&=& \kappa^{-1}_{\rm S}- 
   \frac{g^{}_{\rm S}}{(2\pi)^3}\int_0^1 dy \int d^2 p_\perp 
        \frac{1}{ y (1-y)}
   \left[ 1- \frac{ m_\pi^2 y (1-y)}{m_\pi^2 y(1-y) - (p_\perp^2 + M^2)}
   \right]       \nonumber  \\
  &=&  \left\{ \frac{m_0}{M}+ \frac{2g^{}_{\rm S}}{(2\pi)^3} 
       \int_{0}^{1}\frac{d x}{x} \int d^2 q_\perp \right\}  \nonumber \\
  &&\ - \left[ \frac{ g^{}_{\rm S} }{(2\pi)^3} 
        \int_0^1 \frac{d y}{ y(1-y)}  \int d^2 p_\perp 
             - \frac{ g^{}_{\rm S} }{(2\pi)^3}  
        \int_0^1 dy  \int d^2 p_\perp 
         \frac{ m_\pi^2 }{ m_\pi^2 y (1-y)- (p_\perp^2 + M^2)}\right]\NN
&=&\frac{m_0}{M}+\frac{ g^{}_{\rm S} }{(2\pi)^3}  
       \int_0^1 dy  \int d^2 p_\perp 
       \frac{ m_\pi^2 }{ m_\pi^2 y (1-y)- (p_\perp^2 + M^2)}\, .
  \label{mass_eq_pion}
\end{eqnarray}
In a similar way, the equation for $m_\sigma$ is simplified to  
\BQ
0=\frac{m_0}{M}+\frac{ g^{}_{\rm S} }{(2\pi)^3}  \int_0^1 dy  
\int d^2 p_\perp \frac{ m_\sigma^2 -4M^2 }
                  { m_\sigma^2 y (1-y)- (p_\perp^2 + M^2)}\, .
  \label{mass_eq_sigma}
\EQ
These integral equations (\ref{mass_eq_pion}) and 
(\ref{mass_eq_sigma}) coincide with the results obtained  
 in the previous paper (cf: Eqs.~(4.23) and (4.24) in 
Ref.~\cite{Itakura-Maedan_NJL}). The integral is analytically 
doable with the cutoff indicated in Eq.~(\ref{Eigen_eq_general_regulated}).
Hence, we arrive at the following non-linear equations for the masses 
of scalar and pseudo-scalar mesons:
\BQA
&&\frac{m_0}{M} =\frac{g^{}_{\rm S}\Lambda^2}{4\pi^2}\frac{4M^2}{\Lambda^2}
 r_\pi \left\{
\frac{1}{2}\ln\frac{1+\beta}{1-\beta}-
\sqrt{\frac{1-r_\pi}{r_\pi}}\arctan \frac{\beta}{\sqrt{\frac{1-r_\pi}{r_\pi}}}
\right\}\, ,\label{eigenvalue_pion}\\
&&\frac{m_0}{M} =\frac{g^{}_{\rm S}\Lambda^2}{4\pi^2}\frac{4M^2}{\Lambda^2}
(r_\sigma-1) \left\{
\frac{1}{2}\ln\frac{1+\beta}{1-\beta}-
\sqrt{\frac{1-r_\sigma}{r_\sigma}}\arctan \frac{\beta} 
{\sqrt{\frac{1-r_\sigma}{r_\sigma}}}
\right\},\label{eigenvalue_sigma}
\EQA
where we have introduced (square of) the ratios 
of the meson mass to the threshold mass $2M$:
\BQ
r_\pi=\left(\frac{m_\pi}{2M}\right)^2,\qquad
r_\sigma=\left(\frac{m_\sigma}{2M}\right)^2.
\EQ
Let us first look at the chiral limit $m_0=0$. Then, one finds 
trivial solutions $r_\pi=0$, $r_\sigma=1$. Namely, the pseudo-scalar
meson becomes massless (the Nambu-Goldstone boson), 
while the scalar meson appears as a loosely bound state $m_\sigma=2M$.

\begin{figure}
 \epsfxsize=14cm
 \centerline{\epsffile{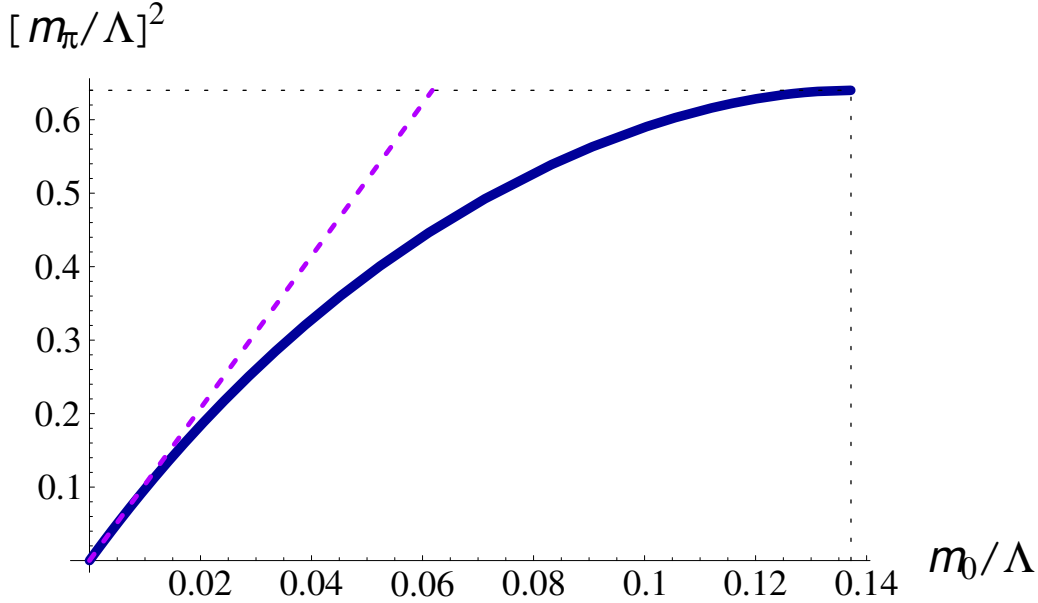}}
 \caption{Pseudo-scalar meson mass squared $(m^{}_\pi/\Lambda)^2$ 
          as a function of $m_0/\Lambda$. The solid line is the numerical 
solution to Eq.~(\ref{eigenvalue_pion}) for $M/\Lambda=0.4$, 
and the dashed line is its approximate solution, Eq.~(\ref{GOR}). 
Two dotted lines are the threshold line $(2M/\Lambda)^2=0.64$ 
(horizontal) and the upper limit of the bound state 
$(m_0/\Lambda)_{\rm upper}=0.137\cdots$ (vertical) as given 
in Eq.~(\ref{upper_bare}). }
 \label{m0dep_pion}
\end{figure}

When $m_0\neq 0$, the equation for the scalar meson does not have a 
(real) solution for $r_\sigma<1$ because the quantity in the curly 
brackets cannot be negative for $r_\sigma<1$. Therefore, the scalar meson 
does not appear as a bound state\footnote{The presentation 
in the previous paper \cite{Itakura-Maedan_NJL} was misleading about 
this point. There, it was argued as if the scalar meson appeared as 
a bound state even for $m_0\neq 0$ case. It does not make sense to 
write $m_\sigma^2=4M^2+{\cal O}(m_0)$ (i.e., Eq.~(4.29) in 
Ref.~\cite{Itakura-Maedan_NJL}) for $m_0\neq 0$ within the present 
approximation.}. This observation is consistent with 
the literature \cite{Sigma,Hatsuda-Kunihiro}. On the other hand, 
the pseudo-scalar meson is still a bound state for small nonzero 
$m_0$ and its mass is estimated as 
\BQ
m_\pi^2=\frac{N}{g^{}_{\rm S}}Z_\pi \frac{m_0}{M} +{\cal O}(m_0^2),
\quad Z_\pi=\frac{1}{N}\left[
\frac{1}{8\pi^2}\ln \frac{1+\beta}{1-\beta}-\frac{\beta}{4\pi^2}
\right]^{-1}.\label{GOR}
\EQ
 In order to compute the decay constant $f_\pi$, one has to determine 
the overall factor $C_\pi$ of the LC wavefunction (\ref{Solution_general}) 
from Eq.~(\ref{normalization}). 
Then, one finds $f_\pi=2M Z_\pi^{-1/2}$, which, combined with Eq.~(\ref{GOR}),
implies the Gell-Mann, Oakes, Renner relation 
$m_\pi^2 f_\pi^2=-4m_0 \langle \bar\Psi\Psi\rangle $, as was 
discussed in Ref.~\cite{Itakura-Maedan_NJL}.
In Fig.~\ref{m0dep_pion}, we show the numerical solution to the 
eigenvalue equation (\ref{eigenvalue_pion}), together with the 
approximate solution (\ref{GOR}). Shown is the square of the mass,
and thus the linear dependence upon $m_0$ is clear for small $m_0$. 
As $m_0$ is increased, the pseudo-scalar meson's mass increases, 
and reaches at the threshold $m_\pi=2M$. The value of bare mass 
$m_0$ which gives 
the threshold mass is easily calculated from Eq.~(\ref{eigenvalue_pion})
where we use $r_\pi=1$ and delete $g^{}_{\rm S}$ dependence
by using the gap equation. Explicitly, the result is 
\BQ
\left(\frac{m_0}{\Lambda}\right)_{\rm upper}={\frac{M}{\Lambda}}
\left\{1+\frac{2-(M/\Lambda)^2(1+\ln\frac{2\Lambda^2}{M^2})}
{2(M/\Lambda)^2 \ln \frac{1+\beta}{1-\beta}}\right\}^{-1}.
\label{upper_bare}
\EQ
For $M/\Lambda=0.4$, this upper bound of the bare mass becomes
$(m_0/\Lambda)_{\rm upper}=0.137\cdots$, which is in agreement 
with the numerical result. 
Therefore, the bound state in the pseudo-scalar channel 
exists even for nonzero bare mass $m_0$
if it is smaller than $(m_0/\Lambda)_{\rm upper}$ given by 
Eq.~(\ref{upper_bare}).

\begin{figure}
 \epsfxsize=14cm
 \centerline{\epsffile{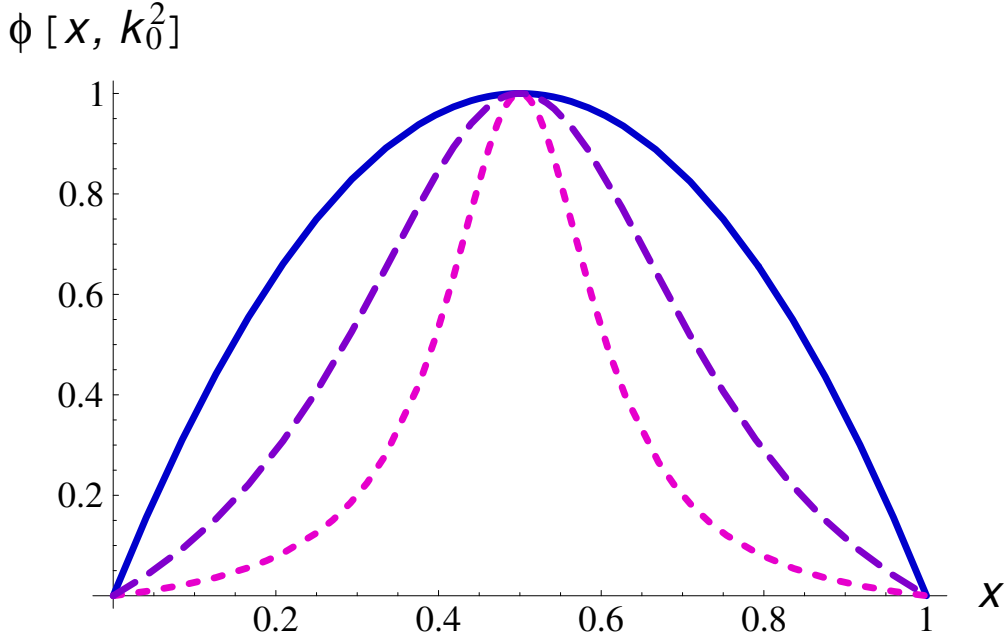}}
 \caption{ The spin-independent LC wavefunctions $\phi(x,\kk^2=0.05 M^2)$ 
of pseudo-scalar (solid), scalar (short-dashed) and 
vector (long-dashed) mesons in the chiral limit $m_0=0$.
Mass of the vector meson is taken to be a typical value 
$m_{\rm V}/2M=0.9$ (see Fig.~\ref{vector_mass}). Rescaled at $x=1/2$ 
just for comparison.}
 \label{LCWF}
\end{figure}

Once we obtain the mass of mesons, we can discuss the shape of the 
spin-independent LC wavefunctions. It is the meson's mass that controls
the shape of wavefunction. Consider the chiral limit, only when
the scalar meson appears as a bound state. Then, the difference of the 
mass between $m_\pi=0$, and $m_\sigma=2M$ greatly affects the shape.
Both have  peaks at $x=1/2$, but widths are significantly different.
The wavefunction of the scalar meson has a very sharp peak at $x=1/2$
(and even divergent for $k_\perp=0$), which clearly shows the constituent
picture. In Fig.~\ref{LCWF}, we plot $\phi(x,k_\perp^2=0.05M^2)$'s of 
scalar and pseudo-scalar mesons in the chiral limit, which are compared 
with that of the vector meson to be derived soon. Transverse 
momentum has been taken nonzero in order to avoid the singular 
behavior at $\kk=0$. The wavefunctions are rescaled at $x=1/2$ 
just for comparison.

\subsection{Vector mesons}
We have been concerned if the bound-state equations for the transverse 
and longitudinal modes (Eqs.~(\ref{EQ:TRANS}) and (\ref{EQ:LONG})) 
are equivalent to each other. 
This can be explicitly shown as follows. 
First of all, we can verify that both the equations
(\ref{EQ:TRANS}) and (\ref{EQ:LONG}) derive the same equation 
for a vector meson mass $m^{}_{\rm V}$ with the extended parity 
invariant cutoff (\ref{LB}). Namely, 
Eq.~(\ref{Eigen_eq_general_regulated}) for the transverse and 
longitudinal potentials
$$
V_{\rm T}= -\frac{g^{}_{\rm V}}{(2\pi)^3}
\frac{p_\perp^2+M^2 -2y(1-y)p_\perp^2}{y^2(1-y)^2}\, , \qquad 
V_{\rm L}=-\frac{g^{}_{\rm V}}{(2\pi)^3}
\frac{4(p_\perp^2+M^2)}{y(1-y)}\, ,
$$
 leads to the same equation.
This, together with the fact that the spin-independent LC wavefunctions 
for transverse and longitudinal modes have the same functional form 
(\ref{Solution_general}), ensures the equivalence of the two 
bound-state equations. If the masses of transverse and 
longitudinal vector mesons coincide with each other, then so do the 
spin-independent LC wavefunctions, as well as the bound-state 
equations\footnote{Notice that we can derive the eigenvalue equations 
simply by inserting the  LC wavefunction (\ref{Solution_general}) 
into the (r.h.s. of the) bound-state equations with the extended parity 
invariant cutoff. 
Since the LC wavefunction (\ref{Solution_general}) is a direct 
consequence of the bound-state equations, to obtain the same 
equation for $m_{\rm V}$ means that the original equations are 
also equivalent to each other.}. 
Indeed, after some calculation, one can see that 
Eq.~(\ref{Eigen_eq_general_regulated}) with $i=$ T and $i=$ L 
reduces to 
the same equation for the vector meson mass $m_{\rm V}$: 
\begin{equation}
   \frac{1}{ \widetilde g^{}_{\rm V}}
   =  \frac{2}{3} \left[ \, \beta  + (1- \beta^2) \left\{
        r^{}_{\rm V} \ln \left( \frac{1+\beta}{1-\beta} \right)   
        - \left( 2  r^{}_{\rm V} +1 \right)  
        \sqrt{\frac{1-r^{}_{\rm V}}{r^{}_{\rm V}} }
        \arctan \frac{\beta} {\sqrt{\frac{1-r^{}_{\rm V}}{r^{}_{\rm V}} } } 
          \right\}  \right],
  \label{eq_for_mV}
\end{equation}
where we have defined a dimensionless coupling constant 
$\widetilde g^{}_{\rm V} =  ({  \Lambda^2}/{ 4 \pi^2})\, g^{}_{\rm V}$ 
and $r^{}_{\rm V}$ similarly as for the scalar and pseudo-scalar cases:
\begin{equation}
   r^{}_{\rm V} \equiv \left( \frac{ m^{}_{\rm V}}{2M} \right)^2.
 \label{ratio}
\end{equation}
This completes the proof of the equivalence between the 
transverse and longitudinal equations.

Let us consider the physics consequences of the eigenvalue equation
(\ref{eq_for_mV}).
A physical bound-state should appear only when the ratio $r^{}_{\rm V}$ is 
in the range $ 0 < r^{}_{\rm V} < 1$. Equation (\ref{eq_for_mV}) 
has a solution in this region when the strength of the coupling 
constant $\widetilde g^{}_{\rm V}$ is in the range 
$\widetilde g_{\rm V}^{\rm{(min)}} < \widetilde g^{}_{\rm V} <  
\widetilde g_{\rm V}^{\rm{(max)}}$ 
defined by
\begin{eqnarray}
    \widetilde g_{\rm V}^{\rm{(min)}}  \equiv
   \frac{3}{2}  \left\{ \beta  + (1- \beta^2)
        \ln \left( \frac{1+\beta}{1-\beta} \right) \right\}^{-1},\quad
   \widetilde g_{\rm V}^{\rm{(max)}}  \equiv 
\frac{3}{2} \cdot \frac{1}{\beta^3 }\, .\label{coupling_limits}
\end{eqnarray}
Two limiting cases $ \widetilde g^{}_{\rm V} = \widetilde g_{\rm V}^{\rm{(min)}} $ and 
$ \widetilde g^{}_{\rm V} =\widetilde g_{\rm V}^{\rm{(max)}} $ 
correspond to $r^{}_{\rm V}=1$ (loose binding limit), and $r^{}_{\rm V}=0$ (tight binding limit), 
respectively. When $M/ \Lambda \rightarrow 0\ (\beta\to 1)$, the physical 
bound-state region shrinks $\widetilde g_{\rm V}^{\rm{(min)} } \rightarrow 
\widetilde g_{\rm V}^{\rm{(max)} } $, while it becomes wider as $M/ \Lambda $ 
grows large. The existence of $\widetilde g_{\rm V}^{\rm (min)}$ is consistent 
with the observation that there is no bound state in the NJL model
without the vector interaction. 
Similar behaviors have been found in Ref.~\cite{Dmitra}.

\begin{figure}
 \epsfxsize=14cm
 \centerline{\epsffile{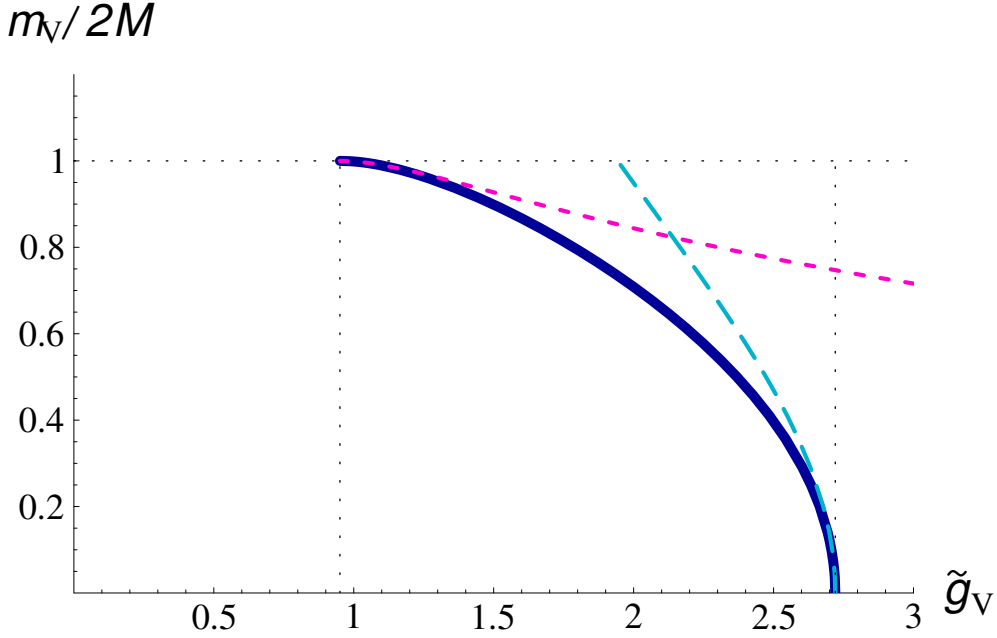}}
 \caption{Vector meson's mass $m_{\rm V}/2M$ as a function of 
$\widetilde g^{}_{\rm V}$. 
The numerical solution to Eq.~(\ref{eq_for_mV}) (solid line) is 
compared with its approximate solutions Eq.~(\ref{app_mV_1})
 (short-dashed) and Eq.~(\ref{app_mV_2}) (long-dashed).
Constituent quark mass is taken to be $M/\Lambda=0.4$ 
(see Fig.~\ref{condensate}). A bound state appears only in 
the regime 
$\widetilde g_{\rm V}^{\rm{(min)}}=0.95 \ <\widetilde g^{}_{\rm V} < 
\widetilde g_{\rm V}^{\rm{(max)}}=2.72$, which are indicated by 
two vertical dotted lines. The horizontal dotted line corresponds to 
the threshold.}   
 \label{vector_mass}
\end{figure}

One can find approximate solutions to Eq.~(\ref{eq_for_mV}) 
when the vector coupling $\widetilde g^{}_{\rm V}$ is close to either 
its minimum $\widetilde g_{\rm V}^{\rm{(min)}}$ or 
maximum $\widetilde g_{\rm V}^{\rm{(max)} }$.
Namely, 
\BQA
r^{}_{\rm V} &\simeq& 1 
   -\left\{\frac{({\widetilde g^{\rm (min)}_{\rm V}})^{-1}
           -({\widetilde g^{}_{\rm V}})^{-1}}{\pi (1-\beta^2)}\right\}^2
\qquad\ {\rm for}\qquad  \widetilde g^{}_{\rm V}\simge 
    \widetilde g^{\rm (min)}_{\rm V}\, ,\label{app_mV_1}\\
r^{}_{\rm V}&\simeq& \frac{({\widetilde g^{}_{\rm V}})^{-1}
          -({\widetilde g^{\rm (max)}_{\rm V}})^{-1}}
           {\frac{2}{3}(1-\beta^2)\left(\ln \frac{1+\beta}{1-\beta}
            -2\beta\right)}
\qquad\qquad {\rm for}\qquad  \widetilde g^{}_{\rm V}\simle 
         \widetilde g^{\rm (max)}_{\rm V}\, .\label{app_mV_2}
\EQA
When we derived the bound-state equations (\ref{EQ:TRANS}) and 
(\ref{EQ:LONG}), we picked up the leading contribution in the 
(integral form of the) eigen-value equation for $m_{\rm V}$ 
(see Appendix~\ref{BSELONG}).  However, the solution to the 
resulting eigen-value equation (\ref{eq_for_mV}) has highly nontrivial
dependence upon $\widetilde g^{}_{\rm V}$. This is evident in the
above approximate solutions to the equation.

In Fig.~\ref{vector_mass}, a numerical solution to Eq.~(\ref{eq_for_mV})
is shown as a function of $\widetilde g^{}_{\rm V}$, where the 
constituent quark mass is taken to be $M/\Lambda=0.4\ (\beta=0.82)$ 
as an example. As we expect, a bound state appears for 
$\widetilde g^{}_{\rm V}$ larger than the critical value 
$\widetilde g^{\rm (min)}_{\rm V}$, 
and as one increases $\widetilde g^{}_{\rm V}$, 
the mass starts to decrease from the 
threshold value $2M$. The value of critical coupling constants 
are exactly the same as the values predicted by the analytic calculation.
When $\beta=0.82$, they are $\widetilde g_{\rm V}^{\rm{(min)}}=0.95$ and 
$ \widetilde g_{\rm V}^{\rm{(max)}}=2.72$, which are indicated as 
two vertical dotted lines in the figure. Besides, we have plotted 
the approximate solutions Eqs.~(\ref{app_mV_1}) and (\ref{app_mV_2})
in the same figure, and they are in good agreement 
in the regimes where each approximation is valid.

\begin{figure}
 \epsfxsize=14cm
 \centerline{\epsffile{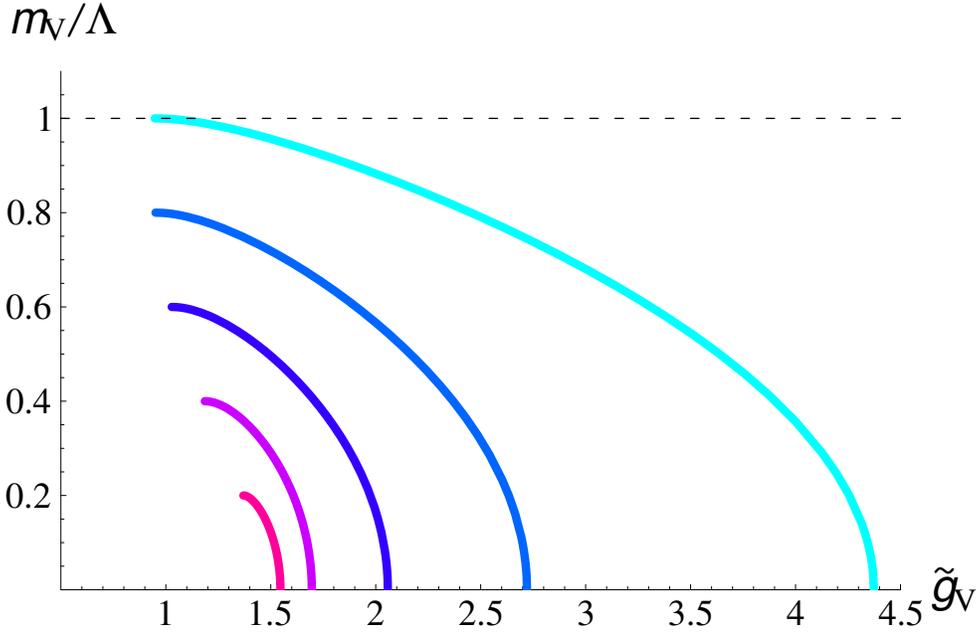}}
 \caption{Vector meson's mass $m^{}_{\rm V}/\Lambda$ as a function of 
 $\widetilde g^{}_{\rm V}$ for different values of constituent mass 
 $M/\Lambda$. From bottom to top, 
 $M/\Lambda=0.1,\, 0.2,\, 0.3,\, 0.4,\, 0.5.$ The line for 
 $M/\Lambda=0.4$ is the same result as shown in Fig.~\ref{vector_mass}. 
 Below  the dashed line is the physical region. These results can 
 be re-interpreted as either $m_0$ or $\widetilde g^{}_{\rm S}$ 
 dependences from 
 Fig.~\ref{condensate}, since the $m_0$ and $\widetilde g^{}_{\rm S}$ 
dependences enter the vector meson's mass only through the constituent 
mass $M$.}
 \label{Mdep_vector}
\end{figure}

It is also interesting to see $m^{}_{\rm V}$ as a function of 
$\widetilde g^{}_{\rm V}$ for different values of the constituent mass.
This is shown in Fig.~\ref{Mdep_vector}.
Unlike in Fig.~\ref{vector_mass}, the vertical axis is
 $m^{}_{\rm V}/\Lambda$
instead of $m^{}_{\rm V}/2M$ since we change the value of constituent 
mass $M/\Lambda=0.1 \sim 0.5$.  
Though the eigenvalue equation (\ref{eq_for_mV}) does not explicitly 
depend upon $\widetilde g^{}_{\rm S}$ and $m_0$, they indirectly affect
the vector meson's mass through the constituent mass $M$ 
(cf. Fig.~\ref{condensate}). Thus, to 
study the $M$ dependence of $m^{}_{\rm V}$ also implies to study
the $\widetilde g^{}_{\rm S}$ or $m_0$ dependence. 
The result in Fig.~\ref{Mdep_vector} is consistent with the 3 dimensional
plot of $m^{}_{\rm V}/\Lambda$ as a function of 
$\widetilde g^{}_{\rm V}$ and $\widetilde g^{}_{\rm S}$
in the chiral limit, which was presented in the letter \cite{PLB}.
As $M/\Lambda$ is increased (i.e., going to larger 
$\widetilde g^{}_{\rm S}$), the bound-state region becomes 
larger as was discussed around  Eq.~(\ref{coupling_limits}).

Lastly, we compare the spin-independent LC wavefunction 
of the vector meson with those of the scalar and pseudo-scalar mesons.
This is shown in Fig.~\ref{LCWF} as already mentioned. 
We plot three wavefunctions 
for nonzero transverse momentum $k_\perp^2=0.05 M^2$ in the chiral 
limit $m_0=0$. Mass of the vector meson is taken to be a typical value 
$m_{\rm V}/2M=0.9$ (see Fig.~\ref{vector_mass}). All the wavefunctions are 
normalized at $x=1/2$ for comparison.
As we discussed before, the difference of shape is 
due to different values of the mass. In the chiral limit, we have 
a relation $0=m_\pi \le m^{}_{\rm V} \le m_\sigma=2M$. 
Thus it is reasonable that the wavefunction of the vector meson lies 
in between those of the scalar and pseudo-scalar mesons. 
The constituent picture works better for a vector meson
than for a pseudo-scalar meson.

\section{Summary and discussions}\label{Summary}
\setcounter{equation}{0}

We have presented a detailed analysis of the application of
light-front quantization to light mesons. Our framework is 
one of the most straightforward approaches towards relativistic 
bound-state systems, and indeed allows us to obtain the light-cone 
wavefunction and mass of the meson simultaneously by solving the 
Hamiltonian eigenvalue equation. One of the subtleties in treating 
light mesons in the light-front formalism is the chiral symmetry 
breaking. However, we know how to describe it at least in the 
Nambu-Jona-Lasinio model as discussed in the present paper. 
All the calculation was done within the NJL model with the vector 
interaction and in the large $N$ limit, where a meson is described 
as a composite state of a quark and an antiquark. 
We mainly focused on the problems of describing vector mesons and 
understanding their properties, though we also reproduced the previous 
results for the scalar and pseudo-scalar mesons in 
Ref.~\cite{Itakura-Maedan_NJL}.
The vector mesons can be formed in the NJL model if one adds the 
vector interaction which strengthens the attractive force between the
 quark and the antiquark. 
For the light mesons, the effect of chiral symmetry breaking is 
of special importance. We followed the same procedures as in 
 Ref.~\cite{Itakura-Maedan_NJL} to incorporate this effect. 
We are able to obtain the LC Hamiltonian in the broken phase,
which then can be used in deriving the bound-state equations. 
In the NJL model, we have to be careful about the regularization 
scheme, but once it is properly done, we can obtain reasonable results
about the LC wavefunctions and masses of light mesons.  
Mass of the vector meson decreases as one increases the strength 
of the vector interaction.
This behavior is consistent with Refs.\cite{Dmitra,Kugo} which also 
treated the vector meson within the NJL model with the vector 
interaction. 

We  utilized several approximations in deriving the bound-state 
equations. For the vector meson, we estimated only the leading 
effect with respect to the vector coupling. In fact, this was enough 
to form a bound state in the vector channel, but somewhat unexpected 
behavior of the mass that it vanishes at some critical vector 
coupling $\widetilde g^{\rm (max)}_{\rm V}$, is probably due to 
this approximation. For the scalar and pseudo-scalar mesons, we
ignored the effects of vector interaction, because bound states 
are possible even with the scalar interaction. We know that the 
vector interaction can modify the structure of pseudo-scalar 
mesons, and it would be important to study this effect. 

 The methods we developed in the present paper seems
specific to the NJL model. This is not true. Actually, the 
applicability of the formalism  is not limited to the NJL model, 
but is much wider. 
For example, the same formalism can be applied to any fermionic models 
without gauge fields, which include models with nonlocal 
current-current interaction generated by either one gluon exchange or 
instantons.

There remain several interesting topics to be considered in the future. 
First, now that we obtain the LC wavefunctions of light mesons, we can 
use them to compute the physical observables such as the distribution 
amplitudes. Of special importance is the physical form factors. 
Such analysis was indeed done for a pseudo-scalar meson \cite{Heinzl}, 
but a similar analysis is possible for the vector meson  \cite{PaperII}. 
Next, our framework can in principle treat the axial vector state in a 
similar way, and it would be interesting to investigate its 
bound-state structure.

\section*{Acknowledgments}
The research of K.I. is supported by the program,  
JSPS Postdoctoral Fellowships for Research Abroad.

\appendix
\setcounter{section}{0}
\section{Notations}\label{Notation}
\setcounter{equation}{0}
Our notation is the 
following: The light-cone coordinates are 
$x^\pm=(x^0\pm x^3)/\sqrt2,$ and $x_\perp^i=(x^1,x^2)=\vec{x}_\perp$, 
which are written together as 
$x^\mu= (x^+,x^-, x_\perp^i)$. 
The derivatives are $\partial_\pm=\partial / \partial x^\pm$ and 
$\del_\perp=\del/\del x_\perp$.
We define the spatial vector $\x$ for coordinates 
by $\x=(x^-,x_\perp^i)$. 
On the other hand, for momenta, we define   
$\Kmbf{p}=(p^+, p^1_\perp, p^2_\perp)$ (notice that $p^+=p_-$ is 
conjugate to $x^-$) and $\p\x=-p^+x^-+p_\perp^i x_\perp^i$ 
so that $px=p^-x^++p^+x^--p_\perp^i x_\perp^i=p^-x^+-\p\x$. 
We use $\mu,\nu$ for Lorentz indices of four vectors, $i,j$ for 
transverse coordinates $1,2,$ and  $\alpha,\beta$ for spinor indices.

The projection operator $\Lambda_\pm$ is defined by 
\BQ
\Lambda_\pm   \equiv  \frac{1}{2} ( 1\pm \gamma^0 \gamma^3 )
\EQ
and satisfies $(\Lambda_\pm)^2=\Lambda_\pm$, and $\Lambda_++\Lambda_-=1$.
Some useful formulas are $\Lambda_- \gamma^+=\gamma^+\Lambda_+$, 
$\Lambda_+ \gamma^-=\gamma^-\Lambda_-$.

\section{LF Quantization of a free fermion: spinor basis}
\label{Free_Fermion}
\setcounter{equation}{0}

Here we summarize the LF quantization of a free fermion
based on the plane wave solutions to the Dirac equation. 
This gives our "spinor basis" which is useful when we treat the
case with vector interaction.

Consider a classical plane wave solution 
$\psi(x) = u(p) \, {\rm e}^{ -i p x}$ to the free Dirac equation:
\begin{equation}
     ( \gamma^\mu p_\mu - M ) u(p) = 0\, ,  \hskip2cm ( \mu = +,-,1,2 )
\end{equation}
where the four momentum $p^\mu$ must satisfy the dispersion relation
\begin{equation}
   p^- = \frac{ {p}_\perp ^2 +M^2 }{2 p^+} .
  \label{dispersion}
\end{equation}
So far, the LC energy $p^-$ can take either positive or negative value, 
but, we redefine the negative energy solution $u(-p^-,-p^+,-p_\perp)$ 
($p^->0,\ p^+>0$) by $v(p^-,p^+,p_\perp)$, and we always treat  $p^-$ to be 
positive. We call $u(\p)$ and $v(\p)$ the positive
and negative energy solutions, respectively. 
It should be noticed that the longitudinal momentum $p^+$ is 
also positive, because the sign of the LC energy and that of the 
longitudinal momentum is correlated as is shown in Eq.~(\ref{dispersion}).
The "positive energy solution" $ u( p^+, {p}_\perp, s ) $ 
satisfies
$0 =  ( \gamma^\mu p_\mu - M ) \, u( p^+, {p}_\perp, s )$
and
${\bar u( p^+, {p}_\perp, s )  } \,  ( \gamma^\mu p_\mu - M ) = 0 ,$
where $s$ is an index to distinguish two different solutions.
Similarly, the "negative energy solution" $v( p^+, {p}_\perp, s )$ 
satisfies
$0 =   ( - \gamma^\mu p_\mu - M ) \ v( p^+, {p}_\perp, s )$ and 
${\bar v( p^+, {p}_\perp, s )  } \,  ( - \gamma^\mu p_\mu - M ) = 0.$ 
Normalization of the solutions are given by 
$$
  {\bar u}( \p, s) \, u( \p,s') = - {\bar v}( \p, s) \, v( \p,s') = 2 \, 
  M\, \delta_{s s'} \, ,
$$
and the followings are useful:
\begin{eqnarray}
   {\bar u}( \p, s) \gamma^\mu \, u( \p,s') &=&  {\bar v}( \p, s)
\gamma^\mu  \, v( \p,s')
       = 2 \, p^\mu \, \delta_{s s'} ,    \nonumber  \\
  \sum_{s} u( \p,s) \, {\bar u}( \p, s) &=& p_\mu \gamma^\mu + M ,
\nonumber  \\
  \sum_{s} v( \p,s) \, {\bar v}( \p, s) &=& p_\mu \gamma^\mu - M .\nonumber
\end{eqnarray}
The fermion field can be expanded by use of these complete solutions,
\begin{eqnarray}
 \psi (x) &=& \frac{1}{\left(2\pi\right)^{3/2}} 
              \int_{0}^{\infty} d p^+ \int_{- \infty}^{\infty} d^2 {p}_\perp  
         \frac{1}{\sqrt{2 p^+}}     \nonumber  \\
  & &   \times
      \sum_{s=\pm} \left[b (p^+, {p}_\perp, s) \ u(p^+, {p}_\perp, s  )  \, 
              {\rm e}^{-i p \cdot x}
      + d^{\dagger} (  p^+, {p}_\perp, s  ) \ v(  p^+, {p}_\perp, s  )
     \, {\rm e}^{i p \cdot x}\right] , 
\label{mode_exp_full}  
\end{eqnarray}
where $b (p^+, {p}_\perp, s)$ and $d^{\dagger} (p^+, {p}_\perp, s)$ 
are c-number coefficients. Defining
$ \psi = \Lambda_+ \psi + \Lambda_- \psi \equiv  \psi_+ + \psi_- , $ 
 the good component $\psi_+$ is expanded as
\begin{eqnarray}
 \psi_+ (x) &=& \frac{1}{\left(2\pi\right)^{3/2}} 
            \int_{0}^{\infty} d p^+ \int_{- \infty}^{\infty} d^2 {p}_\perp  
            \frac{1}{\sqrt{2 p^+}}     \nonumber  \\
  & &   \times
      \sum_{s=\pm} \left[b (p^+, {p}_\perp, s) \ u_+(p^+, {p}_\perp, s)  
         \, {\rm e}^{-i p \cdot x}
      + d^{\dagger} (  p^+, {p}_\perp, s  ) \ v_+(  p^+, {p}_\perp, s  )
     \, {\rm e}^{i p \cdot x}\right] ,
   \label{mode_exp_classical}
\end{eqnarray}
where $u_+=\Lambda_+ u$ and $v_+=\Lambda_+ v$. These projected 
spinors satisfy the following relations:
\begin{eqnarray}
u_+^{\dagger}( \p,s') u_+( \p,s) &=&
v_+^{\dagger}( \p,s') v_+( \p,s) = \sqrt{2} p^+\,\delta_{ss'} \NN
\sum_s u_+( \p,s) u_+^{\dagger}( \p,s) &=&
\sum_s v_+( \p,s) v_+^{\dagger}( \p,s) = \sqrt{2} \Lambda_+ p^+
\,\, . \nonumber
\end{eqnarray}

Since $ \psi_- $ is a dependent variable, 
we impose the quantization condition only on the good component 
$\psi_+$ (see Eqs.~(\ref{quantization1}), (\ref{quantization2})). 
The mode expansion for  $\psi_+$ is given in the same form as 
Eq.~(\ref{mode_exp_classical}), but now 
$b (  p^+, {p}_\perp, s  )$ and 
$d^{\dagger} (  p^+, {p}_\perp, s  )$ are operators.

The mode expansion of the bad component $\psi_-$ should be
 given through the constraint equation 
$ \psi_-(x) = \frac{1}{ 2 i \partial_- }  
( i \gamma_{\perp}^i \partial_i + M ) \gamma^+ \psi_+$ by using 
Eq.~(\ref{mode_exp_classical}). 
One can explicitly show that $\psi=\psi_++\psi_-$ reproduces the 
mode expansion Eq.~(\ref{mode_exp_full}).

\section{Boson expansion method as the $1/N$ expansion of ${\sf M}(\p,\q)$}
\label{BEM}
\setcounter{equation}{0}

In the boson expansion method, we rewrite the bilocal operators 
${\sf M}(\p,\q)$ defined by Eq.~(\ref{bilocal_M}) with respect to
bosonic operators so that the commutation 
relation between ${\sf M}(\p,\q)$'s is correctly reproduced.
Physically this corresponds to capturing a composite state made of 
a fermion and an antifermion, as a bosonic state with additional 
"structure" due to statistics. Especially, when $N$ is large, 
the composite state behaves as a boson, and thus large $N$ expansion
of the bilocal operator is naturally realized by the boson 
expansion method (of the Holstein-Primakoff type).

The bilocal operator ${\sf M}(\p,\q)$   satisfies the following
commutator:
\begin{eqnarray}
 &&\hspace{-2cm}\Big[: {\sf M}_{\alpha_1 \alpha_2 }(\p_1,\p_2):\, ,\ \
               :  {\sf M}_{\beta_1 \beta_2 }(\q _1,\q _2):\Big]\\
 &&\hspace{-1cm}= \, :  {\sf M}_{\alpha_1 \beta_2 }(\p_1,\q _2):
(\Lambda_+)_{\alpha_2 \beta_1 }\delta (\p_2+\q _1)
    \, -\, :  {\sf M}_{\beta_1 \alpha_2 }(\q _1,\p_2):
(\Lambda_+)_{\beta_2 \alpha_1  }
                                     \delta ( \q _2 + \p_1 )
\nonumber \\
 & &\hspace{-1cm}\quad +\, N(\Lambda_+)_{\beta_2 \alpha_1} \delta(\q_2 + \p_1)
          (\Lambda_+)_{\alpha_2 \beta_1 }
       \delta (\p_2+\q _1)  \nonumber \\
 &&   \times 
\Big(
\theta(p_{1}^+)\theta(p_{2}^+)\theta(-q_{1}^+)\theta(-q_{2}^+)-
      \theta(-p_{1}^+)\theta(-p_{2}^+) \theta(q_{1}^+) \theta(q_{2}^+)
\Big),  \nonumber
\end{eqnarray}
and its vacuum expectation value is 
\begin{equation}
 \langle 0 \vert {\sf M}_{\alpha\beta}(\p,\q ) \vert 0 \rangle
  = N \theta(p^+) \delta ( \p+\q  ) (\Lambda_+)_{\beta \alpha} .
\end{equation}
This complicated commutator can be reproduced by replacing ${\sf M}(\p,\q)$
by some simple functions of a (bilocal) bosonic operator.

Let us first introduce bilocal bosonic operators $B_{\alpha \beta}(\p,\q)$
which have  Dirac indices. We impose  they satisfy 
the following commutation relations:
\BQA
 &&  [ B_{\alpha_1 \alpha_2}(\p_1,\p_2),\, B_{\beta_1
\beta_2}^\dagger(\q _1,\q _2)]
   =(\Lambda_+)_{\alpha_1 \beta_1 } \delta(\p_1-\q _1)
    (\Lambda_+)_{\beta_2 \alpha_2} \delta(\q _2 - \p_2),\\
 &&[ {B}_{\alpha_1 \alpha_2} (\p_1,\p_2),  {B}_{\beta_1 \beta_2} (\q _1,\q _2) ]  =  0  \qquad 
  (p_i^+,q_i^+ > 0 ), 
\EQA
and the following properties
\begin{eqnarray}
&& (\Lambda_+)_{\alpha_1 \gamma_1} B_{\gamma_1 \gamma_2 }(\p_1, \p_2)
        =  B_{\alpha_1 \gamma_2 }( \p_1, \p_2) , \hskip 0.5cm
   B_{\gamma_1 \gamma_2 }(\p_1, \p_2) (\Lambda_+)_{\gamma_2 \alpha_2}
        = B_{\gamma_1 \alpha_2 }( \p_1, \p_2)\, , \\
&& (\Lambda_+)_{\gamma_1 \alpha_1} B^{\dagger}_{\gamma_1 \gamma_2}(\p_1, \p_2)
        =  B^{\dagger}_{\alpha_1 \gamma_2 }( \p_1, \p_2) , \hskip 0.5cm
   B^{\dagger}_{\gamma_1 \gamma_2 }( \p_1, \p_2)(\Lambda_+)_{\alpha_2 \gamma_2}
        = B^{\dagger}_{\gamma_1 \alpha_2 }( \p_1, \p_2)\, ,
\end{eqnarray}
and so on. Then the  operators $: {\sf M} :$ can be represented as 
\begin{eqnarray}
  :  {\sf M}^{-+}_{\alpha_1 \alpha_2}(\p_1,\p_2):
  &=& \int_0^\infty dq^+\int_{-\infty}^\infty d^2q_\perp
      B^\dagger_{\alpha_1 \gamma} (-\p_1,\q )
      B_{\alpha_2 \gamma}( \p_2, \q )\nonumber \\
  &\equiv & {\cal A}_{\alpha_2 \alpha_1}(\p_2,-\p_1),\\
  :  {\sf M}^{+-}_{\alpha_1 \alpha_2}(\p_1,\p_2):
  &=& - \int_0^\infty dq^+\int_{-\infty}^\infty d^2q_\perp
     B^\dagger_{\gamma \alpha_2} (\q , -\p_2)
      B_{\gamma \alpha_1}( \q , \p_1), \\
  :  {\sf M}^{++}_{\alpha_1 \alpha_2}(\p_1,\p_2):
  &=& \int_0^\infty dq^+\int_{-\infty}^\infty d^2q_\perp
     (\sqrt{N-{\cal A}})_{\alpha_2 \gamma}(\p_2,\q )
       B_{\gamma \alpha_1}(\q ,\p_1),   \\
  :  {\sf M}^{--}_{\alpha_1 \alpha_2}(\p_1,\p_2):
  &=& \int_0^\infty dq^+\int_{-\infty}^\infty d^2q_\perp
      B^\dagger_{\gamma \alpha_2} (\q ,-\p_2)
       (\sqrt{N-{\cal A}})_{\gamma \alpha_1}(\q ,-\p_1),
\end{eqnarray}
where the upper indices stand for the signs of the longitudinal momenta.
This is the boson expansion method of the Holstein-Primakoff type, 
represented in the "Dirac basis" (namely, the bosonic operators explicitly
have the Dirac indices). 
This is essentially the same as the one used in 
Ref.~\cite{Itakura-Maedan_NJL}. 
Note that the bosonic operator is of ${\cal O}(N^0)$, and there is 
explicit dependence on $N$ on the right-hand sides. 
Thus, one can expand the right-hand sides with respect to $1/\sqrt{N}$,
yielding Eq.~(\ref{M_expansion}) in the text. It should be noticed that 
the natural expansion parameter is $1/\sqrt{N}$ instead of $1/N$
because we have expanded $\sqrt{1-{\cal A}/N}$.
The first few terms are given as follows:
\begin{eqnarray}
  {\sf m}_{\alpha\beta}^{(0)}(\p,{\q})
    &=& ( \Lambda_+ )_{\beta \alpha} \delta (\p+{\q})
\theta(p^+) \theta(-q^+)~,
        \label{HP_lowest}\\
 {\sf m}_{\alpha\beta}^{(1)}(\p,{\q})
    &=& B_{\beta\alpha}({\q},\p)\theta(p^+)\theta(q^+)
       + B_{\alpha\beta}^\dagger (-\p, -\q)\theta(-p^+)\theta(-q^+)~,  
          \label{HP_next} \\
{\sf m}_{\alpha\beta}^{(2)}(\p,\q)
    &=& \int[d \k ] \, 
         B^\dagger_{\alpha\gamma}(-\p, \k  ) 
 B_{\beta\gamma}(\q, \k  )
         \theta(-p^+)\theta(q^+)                   \nonumber \\
      &-& \int[d \k  ]  \, 
         B^\dagger_{\gamma\beta}(\k ,  -\q  )
B_{\gamma\alpha}(\k , \p )
         \theta(p^+)\theta(-q^+)~.    \label{HP_next_to_next}
\end{eqnarray}
The leading term ${\sf m}_{\alpha\beta}^{(0)}(\p,\q)$ comes from
the vacuum expectation value of ${\sf M}(\p,\q)$. The result
${\sf m}_{\alpha\beta}^{(1)}(\p,{\q})$ means that the bilocal 
operator ${\sf M}(\p,\q)$ can be treated as a bosonic operator 
in the first nontrivial leading order.

One can further introduce bilocal bosonic operators 
$ B(\p_1, s_1 : \p_2 , s_2 )$ in the "spinor basis", 
namely, with spins instead of simple Dirac indices.
The relation between operators in the "Dirac basis" and in the 
"spinor basis" is given by 
\begin{eqnarray}
&&   B_{\alpha_1 \alpha_2}(\p_1,\p_2)
   = \frac{1}{\sqrt{2 p_1^+ p_2^+ } }  \sum_{s_1=\pm}  \sum_{s_2=\pm}  u_{+
\alpha_1}( \p_1,s_1) v^\dag_{+ \alpha_2}( \p_2,s_2)
      B(\p_1, s_1 : \p_2 , s_2 )\, ,
\end{eqnarray}
where the good components of the spinor 
$u_+ = \Lambda_+ u$ and  $v_+ = \Lambda_+ v$ are
introduced in Appendix~\ref{Free_Fermion}.
The operator $B(\p_1, s_1 : \p_2 , s_2 )$ satisfies a very simple
commutator (as shown in the text):
$$
  [B(\p_1,s_1:\p_2,s_2 ), B^\dagger(\p'_1,s_1^{'} : \p'_2,s_2^{'} ) ]
   = \delta_{s_1 s_1^{'} }  \delta (\p_1-\p'_1)
     \delta_{s_2 s_2^{'} }  \delta (\p_2-\p'_2)\, .
$$

\section{Higher order solutions to the fermionic constraint and 
the LC Hamiltonian in the case without vector interaction}\label{Hamiltonian}
\setcounter{equation}{0}
\subsection{Higher order solutions}
The bilocal fermionic constraints (\ref{FC_mom_S}) and (\ref{FC_mom_P})
for the higher orders can be cast into a compact form similar to that 
of the lowest order:
\begin{equation}
\pmatrix{{\sf s}^{(n)}(\p,\q)\cr
{\sf p}^{(n)}(\p,\q)}
= 
\pmatrix{F^{(n)}(\p,\q)\cr
G^{(n)}(\p,\q)}
-g^{}_{\rm S}\frac{\epsilon(p^+)}{q^+}
\int_{-\infty}^{\infty}  \frac{d^3\k}{(2\pi)^3}
\pmatrix{{\sf s}^{(n)}(\k,\p+\q-\k)\cr
{\sf p}^{(n)}(\k,\p+\q-\k)}\, ,\label{FC_higher_order}
\end{equation}
where $F^{(n)}$ and $G^{(n)}$ are known functions 
expressed by ${\sf m}^{(n)}$ and lower order quantities 
${\sf s}^{(m)}$, ${\sf p}^{(m)}$ and ${\sf m}^{(m)}$ with $m<n$. 
For example, in the next leading order ${\cal O}(\sqrt{N})$ with $n=1$:
\begin{eqnarray}\label{F1G1}
\pmatrix{ F^{(1)}(\p,\q)\cr
G^{(1)}(\p,\q) }
   =  \left(\pmatrix{1 \cr
-i\gamma_5}\frac{ ( \gamma^i q_\perp^i + M)}{2q^+}\right)_{\alpha\beta}
{\sf m}^{(1)}_{\alpha\beta}(\p,\q)
      -
\left(\frac{ ( \gamma^i q_\perp^i + M)}{2q^+} \pmatrix{1\cr i\gamma_5}
\right)_{\alpha\beta} {\sf m}^{(1)}_{\alpha\beta}(\q,\p) ,
\end{eqnarray}
and in the next-to-next leading order ${\cal O}(N^0)$ with $n=2$, 
one finds 
\begin{eqnarray}
F^{(2)}(\p,\q) 
&=&\frac{(\gamma^i q_\perp^i+ M)_{\alpha\beta}}{2q^+} 
        \left\{ {\sf m}^{(2)}_{\alpha\beta}(\p,\q) - {\sf m}^{(2)}_{\alpha\beta}(\q,\p)
        \right\}  \nonumber  \\
   & & \hskip0.5cm    -\frac{g^{}_{\rm S}}{2 q^+ }\int 
                 \frac{d^3 \k d^3 {\bl} }{(2\pi)^3}
     \left\{  {\sf m}^{(1)}_{\alpha\beta}(\p, \q -\k -\bl)\, 
         \left[ {\sf s}^{(1)} 
            + i \gamma_5 ~ {\sf p}^{(1)}
         \right]_{\alpha\beta} (\k ,\bl)\right.  \nonumber \\
   && \hskip2cm   \left.
       - \left[  {\sf s}^{(1)} 
            - i \gamma_5 ~ {\sf p}^{(1)} \right]_{\alpha\beta}(\k ,\bl)\, 
              {\sf m}^{(1)}_{\alpha\beta}( \q -\k -\bl, \p )
                \right\} ~,
\end{eqnarray}
and similarly for $G^{(2)}$.

The integral equation (\ref{FC_higher_order}) has a very simple 
structure. One can solve it by integrating the equation so that 
the integral on the left-hand side becomes the same as the 
second term in (\ref{FC_higher_order}).
Namely, the solution is
\begin{eqnarray}
 \pmatrix{{\sf s}^{(n)}(\p,\q)\cr
{\sf p}^{(n)}(\p,\q)} =  \pmatrix{F^{(n)}(\p,\q)\cr G^{(n)}(\p,\q)}
-g^{}_{\rm S}\kappa^{}_{\rm S}\frac{\epsilon(p^+)}{q^+}
   \int_{-\infty}^{\infty} \frac{d^3\k}{(2\pi)^3}
\pmatrix{F^{(n)}(\k ,\p+\q-\k)\cr G^{(n)}(\k,\p+\q-\k)}\ ,
\end{eqnarray}
where $\kappa^{}_{\rm S}$ is a constant defined in Eq.~(\ref{kappa_S}).
Since $F^{(n)}$ and $G^{(n)}$ are expressed by ${\sf m}^{(n)}$ 
and lower order quantities, 
we can in principle write down the solutions for any $n$,
order by order.

\subsection{Deriving the LC Hamiltonian}
The LC Hamiltonian is the evolution operator in the LC time ($x^+$) 
direction. In deriving the LC Hamiltonian, we need to follow Dirac's 
procedure for constrained systems regarding $x^+$ as time, since, 
as we already saw, we have the fermionic constraints on the light front.

Then, one finds the following $hermitian$ Hamiltonian:
\begin{eqnarray}
  P^- 
  &=& -\frac{1}{2}\int d^3 \x  \left[ \bar \Psi  i \gamma^i \partial_i ~
\Lambda_- \Psi
          - ( \partial_i \bar \Psi )  i \gamma^i  ~ \Lambda_+ \Psi - m_0
\bar \Psi \Psi \right] \, ,
  \label{Ham_wo_Vector}
\end{eqnarray}
where $ i=1,2 $ and  we have used the fermionic constraint 
(\ref{FCwoVector}) since the model is of the second class 
(in Dirac's terminology). 
If one takes its face value, this Hamiltonian, 
being independent of the coupling constant $G_{\rm S}$, looks equivalent 
to the free Hamiltonian. However, if one rewrites it 
with respect to the physical degree of freedom, -- namely the good 
component of the spinor --, then one obtains a non-trivial 
Hamiltonian that depends explicitly on the coupling constant. 
This is of course because the bad component 
$\psi_- $ is subject to the fermionic constraint (\ref{FCwoVector})
which carries the information of interaction.

Since the Hamiltonian (\ref{Ham_wo_Vector}) is bilinear with respect to
the spinor, one should be able to rewrite it by the bilocal 
operators. This is indeed the case if one introduces a new bilocal 
operator ${\sf V}^i_{\rm I}$ (see Eqs.~(\ref{Ham_bilocal}) and 
(\ref{def_VI})). 
As mentioned in the text, the bilocal operator ${\sf V}^i_{\rm I}$ 
can be related to the known ones, ${\sf M,\, S,\, P}$, through 
an equation which is easily derived from the fermionic constraint. 
In momentum space, it is given by 
\begin{eqnarray}
 i q^+ {\sf V}^i_{\rm I} (\p,\q ) &=&
    - \, \frac{1}{2}  \left\{  \gamma^i_\perp 
    ( \gamma^j_\perp q_\perp^j + m_0 )\right\}_{\alpha\beta}
                  {\sf M}_{\alpha\beta}(\p,\q )
+ \, \frac{1}{2}  \left\{ (  \gamma^j_\perp q_\perp^j + m_0 )
\gamma^i_\perp  \right\}_{\alpha\beta}
                  {\sf M}_{\alpha\beta}(\q ,\p)
\nonumber \\
   && +\, \frac{G_{\rm S}}{2}\int \frac{d^3 \k\, d^3 \bl}{(2\pi)^3}
     \left[ {\sf M}_{\alpha\beta}(\p, \q -\k -\bl)\, 
         \left\{ \gamma^i_\perp \Big(  {\sf S}_{\rm R} 
            +  i \gamma_5 \, {\sf P}_{\rm R} \Big)
\right\}_{\alpha\beta} (\k ,\bl) \right.  \nonumber \\
   && \hskip1cm   \left.
       - \left\{ \Big(  {\sf S}_{\rm R} 
            -  i \gamma_5 \, {\sf P}_{\rm R}  \Big)
\gamma^i_\perp \right\} _{\alpha\beta}(\k ,\bl)\, 
              {\sf M}_{\alpha\beta}( \q -\k -\bl, \p )
                \right] \, .  
\end{eqnarray}
Therefore, given the solutions ${\sf s}^{(n)}_{\rm R}$ and 
${\sf p}^{(n)}_{\rm R}$, one can 
immediately obtain ${\sf v}_{\rm I}^{i\, (n)}$, the expansion coefficients of
${\sf V}_{\rm I}^{i}$.
Hence, one can compute the LC Hamiltonian order by order.
\begin{eqnarray}
  H &=& N\sum_{n=0}^\infty\left(\frac{1}{\sqrt{N}}\right)^n
                h^{(n)}_{\rm S}~,\NN
h^{(n)}_{\rm S}&=&\frac{1}{2} \int d^3 \p\, ( -i p^i_\perp) ~ 
{\sf v}^{i ~(n)}_{\rm I} ( \p, -\p)
         +   \frac{m_0}{2} \int d^3 \p  ~{\sf s}^{(n)}_{\rm R} (  \p, -\p)\, .
\nonumber
\end{eqnarray}
We substitute the "broken" solutions of ${\sf s}^{(n)}_{\rm R}$ and 
${\sf p}^{(n)}_{\rm R}$ to obtain the Hamiltonian which describes 
the broken phase of the chiral symmetry. It turns out that the 
lowest order $ h^{(0)}_{\rm S}$ 
is divergent but is just a constant, and thus can be neglected for the
present purpose. 
The next order ${\sf v}_{\rm I}^{ i ~ (1) }  (\p,{-\p}) $ and
${\sf s}^{ (1) }_{\rm R}  (\p,{-\p}) $ become strictly zero, and 
we have $ h^{ (1) }_{\rm S} =0 $. 
Therefore, the first nontrivial contribution is given by the 
next-to-leading order $n=2$.
After straightforward calculation, one arrives at 
the following expression:
\begin{eqnarray}
  h^{(2)}_{\rm S} 
  & = &  - \int d^3 \p ~ \frac{ p_\perp^2 + M^2 }{2 p^+ }
~{\sf m}^{(2)}_{\alpha \alpha}  (  \p, -\p)   \nonumber  \\
  & &  \hskip0.2cm   - \frac{g^{}_{\rm S}\kappa^{}_{\rm S}}{2}
          \int_{-\infty}^{\infty} 
\frac{d^3 \p\, d^3 \q\, d^3 \k\, d^3 \bl}{(2\pi)^3}
       \delta( \k +  \bl+ \p  +\q ) 
\nonumber  \\
  & &  \hskip2cm   \times \left\{ F^{(1)} ( \p , \q ) F^{(1)} ( \k , \bl  ) 
+ G^{(1)} ( \p , \q ) G^{(1)} ( \k , \bl)
       \right\} \, ,    \nonumber
\end{eqnarray}
where we have again ignored a c-number contribution, which is 
unimportant for our present purpose. 
It is convenient to express the functions $F^{(1)}$
  and  $G^{(1)}$  given in Eq.~(\ref{F1G1}) by using the spinors 
(below, $p^+ >0,\ q^+ >0$):
\begin{eqnarray}
 &&\hspace*{-5mm}
F^{(1)}( -\p, -\q) +  F^{(1)}( -\q, -\p )    
   = \frac{1}{ \sqrt{ 2 p^+ q^+ } } \sum_{s_1, s_2}
             \frac{1}{\sqrt{2} } ~ \{ {\bar u} ( \p,s_1)  v( \q,s_2) \}
B^{\dag} ( \p, s_1 : \q, s_2 )
         + ( \p \leftrightarrow \q ),          \nonumber  \\
 &&\hspace*{-5mm}G^{(1)}( -\p, -\q) +  G^{(1)}( -\q, -\p )   
   = \frac{1}{ \sqrt{ 2 p^+ q^+ } } \sum_{s_1, s_2}
             \frac{1}{\sqrt{2} } ~ \{ {\bar u} ( \p,s_1)  ( i \gamma_5 )
v( \q,s_2) \} B^{\dag} ( \p, s_1 : \q, s_2 )
         + ( \p \leftrightarrow \q )\, .    \nonumber
\end{eqnarray}
Using these expression, together with 
the explicit representation for ${\sf m}^{(1,2)}(\p,\q)$
 (cf. Eqs.~(\ref{HP_next}), (\ref{HP_next_to_next})), 
one can rewrite $h^{(2)}_{\rm S}$ with respect to the bilocal 
boson operators to obtain the final expression (\ref{h2_wo_Vector}).

\section{Meson states with nonzero transverse momentum $P_\perp\neq 0$}
\label{LCWF_NonzeroP}
\setcounter{equation}{0}

Here we present the case with nonzero transverse momentum 
$P_\perp\neq 0$ which requires only a small generalization of the case 
with $P_\perp=0$. In fact, this is almost trivial from the viewpoint 
of the light-front formalism and even not necessary for the present 
purpose of this paper. However, it is useful to have a
formalism with nonzero transverse momentum when we will compute the 
form factors of mesons \cite{PaperII}.  

A generic meson state with nonzero transverse momentum $P_\perp\neq 0$ 
is given by
\begin{eqnarray}
\vert~{\rm meson}~;  {\PP}\, \rangle 
 & = & P^+  \int_{0}^{1} dx_1 dx_2  \, \delta \, ( 1- \sum_{i=1}^{2} x_i \, )
         \int_{-\infty}^{\infty} d^2 k_{1 \perp } ~d^2 k_{2 \perp } \,
       \delta \, ( \, \sum_{i=1}^{2} k_{i \perp }  \, )     \nonumber \\
  & & \hskip0.5cm  \times \sum_{s_1,s_2} \Phi ( x_i, k_{i \perp}, s_1, s_2 )
                 B^\dag ( \p_1, s_1 : \p_2, s_2 )  ~\vert 0 \rangle\, ,
\end{eqnarray}
where $\p_1$ and $\p_2$ are the momenta of a quark and an antiquark
$\PP=\p_1 + \p_2$, and the relations to the "relative" momentum 
coordinates $x_i$ and $k_{i \perp }$ are given by 
$\p_i =( x_i P^+, x_i P_\perp + k_{i \perp} )$.
These variables $x_i$ and $k_{i \perp}$ are invariant under any 
boost transformations. If one performs $x_2$ and $k_{2\perp}$
integrations, one obtains
\begin{equation}
\vert~{\rm meson}~;  {\PP}\, \rangle 
 =  P^+  \int_{0}^{1} dx_1 
         \int_{-\infty}^{\infty} d^2 k_{1 \perp } 
         \sum_{s_1,s_2} \Phi  ( x_1, k_{1 \perp}, s_1,s_2 )
                 B^\dag ( \p_1, s_1 : \PP-\p_1, s_2 )  ~\vert 0 \rangle\, ,
\end{equation}

As was done in the case with $P_\perp=0$, 
the kinematical structure of the LC wavefunction 
$\Phi ( x_1, k_{1 \perp}, s_i )$ is determined through the interpolating 
field of each meson. For example, the LC wavefunctions 
of the scalar and pseudo-scalar mesons are written as follows:
\begin{eqnarray}
    \Phi_\pi ( x_1, \k_{1 \perp},  s_i )
    & \equiv &
     \phi_\pi (x_1, \k_{1 \perp }  ) \,  \frac{1}{ 2 \sqrt{  x_1 (1-x_1)} } \,
     \Big\{ {\bar u} ( \p_1,s_1) \, i \gamma_5 \, v( \PP-\p_1, s_2) \Big\}\, ,
      \nonumber\\   
    \Phi_\sigma ( x_1, \k_{1 \perp},  s_i )
    & \equiv &
     \phi_\sigma (x_1, \k_{1 \perp }  ) \,  \frac{1}{ 2 \sqrt{x_1 (1-x_1)} } \,
     \Big\{ {\bar u} ( \p_1,s_1)   v( \PP- \p_1, s_2) \Big\}\, .\nonumber   
\end{eqnarray}
In fact, one can explicitly show that the right-hand sides are 
independent of the total momentum $\PP$. Therefore, the resulting 
bound-state equations for the spin-independent part of the LC wavefunctions
are exactly the same as those derived in the frame with $P_\perp=0$.

What is less trivial is the case of the vector meson 
due to the projection onto the three physical modes. 
 From the interpolating field, one obtains the expected result:
\begin{eqnarray}
    \Phi_\rho^\lambda(x_1, \k_{1 \perp},  s_i )
    & \equiv &
           \phi_\rho (x_1, \k_{1 \perp }) \,  
            \frac{1}{ 2 \sqrt{  x_1 (1-x_1)} } \,
            \epsilon_\mu ( \lambda, P )   \Big\{ {\bar u} ( \p_1,s_1) \,
\gamma^\mu \, v( \PP - \p_1, s_2) \Big\}\, ,
\end{eqnarray}
but the polarization vector in a generic frame becomes more complicated
\cite{Form_Factor}
\begin{eqnarray}
    \epsilon^\mu ( \lambda=\pm 1, P ) &=&
         ( \, 0, \frac{ \mp P_x-i P_y }{ \sqrt{2} P^+ }  ,  \frac{\mp 1}{
\sqrt{2} } , \frac{-i}{ \sqrt{2} } \, ) \, ,       \nonumber \\
    \epsilon^\mu ( \lambda=0, P ) &=&
         (\frac{P^+}{ m^{}_{\rm V} }, 
          \frac{ P_\perp^2 - m^2_{\rm V} }{ 2 \, m^{}_{\rm V}\, P^+ },
                \frac{P_x}{m^{}_{\rm V}},  \frac{P_y}{m^{}_{\rm V}} )\, ,
\nonumber
\end{eqnarray}
satisfying $ \epsilon_\mu (\lambda, P) \epsilon^\mu (\lambda', P)^*
=-\delta_{\lambda \lambda'},~
                    P_\mu \epsilon^\mu (\lambda, P) =0$. Of course this vector 
reproduces Eq.~(\ref{polarization}) when $P_x=P_y=0$.
It is very important to recognize that the spin-dependent part 
of the LC wavefunction, 
$\epsilon_\mu ( \lambda, P ) ~
      {\bar u} ( \p_1,s_1) \,
\gamma^\mu \,
         v( \p_2 , s_2) $ 
with $\p_1= ( x_1 P^+, x_1 {P}_\perp + k_{1 \perp} )$ and 
$\p_2=( x_2 P^+, x_2 {P}_\perp + k_{2 \perp} )$, 
  does not depend on the total momentum $\PP$. 
This can be explicitly verified as follows. 
Using the Kogut-Super convention, one finds for $\lambda=+1$
\begin{eqnarray}
   & & \hskip-1cm   \epsilon_\mu ( +1, P ) ~
  {\bar u} (  ( x_1 P^+, x_1 {P}_\perp + k_{1 \perp} ),s_1) \, \gamma^\mu \,
         v( ( x_2 P^+, x_2 {P}_\perp + k_{2 \perp} ) , s_2)
\nonumber  \\
   &=&  \frac{ \sqrt{2} }{\sqrt{ x_1 x_2 } }
     \left(  \begin{array}{cc}
        - M        &     x_1 ( k_{2 \perp}^x + i k_{2 \perp}^y )          \\
       x_2 ( k_{1 \perp}^x + i k_{1 \perp}^y )           &     0
                  \end{array}
                    \right)_{s_1  s_2}   \ ,\nonumber
\end{eqnarray}
and similarly for other polarizations. 
Therefore, from these explicit calculations, one concludes that the 
bound-state equations (for the relative motion of a quark and an antiquark) 
in the frame with nonzero transverse momentum 
are the same as those with $P_\perp=0$.

\section{Bound-state equation of longitudinally polarized vector meson}
\label{BSELONG}
\setcounter{equation}{0}

In this Appendix, we first derive the potential (\ref{EQ:STLONG}) 
for the longitudinally polarized vector meson and then 
the bound-state equation (\ref{EQ:LONG}) after taking the leading contribution
with respect to the coupling constant $g^{}_{\rm V}$.

The bound-state equation for the longitudinal vector meson is 
derived from Eq.~(\ref{LFBSE}) where the Hamiltonian and the state are 
given by Eq.~(\ref{Hamiltonian_full}) and  Eqs.~(\ref{meson_generic}),
(\ref{Vec_state}), (\ref{Long_state}), respectively. Explicitly, for 
the longitudinal mode (we use $m^{}_{\rm L}$ for the vector meson's mass),
\begin{eqnarray}
   \frac{m_{\rm L}^2}{2P^+} \,  \left\{
         \frac{-m^{}_{\rm L}}{ 2 P^+ } \, \vert \, \rho^+\,\rangle
       + \frac{P^+}{ m^{}_{\rm L} } \, \vert \, \rho^-\, \rangle \right\}
 =  h^{(2)}  \left\{
         \frac{-m^{}_{\rm L}}{ 2 P^+ } \, \vert \, \rho^+\,\rangle
  + \frac{P^+}{ m^{}_{\rm L} } \, \vert \, \rho^-\, \rangle \right\}\, ,
 \nonumber
\end{eqnarray}
where we have chosen the frame $\PP=(P^+,P_\perp)=(P^+,0_\perp)$.
The following relation is useful when we evaluate this equation:
$$
  {\bar u}( \k, s_1 ) \gamma^-   v( \PP -\k, s_2 )
    =  - \frac{1}{ (P^+)^2 } \frac{ k_\perp ^2 +M^2 }{ 2 x(1-x) } \,
  {\bar u}( \k, s_1 ) \gamma^+  v( \PP  -\k, s_2 ).
$$
Then, a straightforward calculation yields the following 
bound-state equation:
\begin{eqnarray}
  & & \hspace*{-1cm} m_{\rm L}^2 \, \left\{ 1+ \frac{1}{m_{\rm L}^2} \,
                  \frac{ k_\perp ^2 +M^2 }{ x(1-x) } 
         \right\} \phi^{}_{\rm L} (x, k_\perp )  \NN
  &=&  \left\{ 1+  \frac{1}{m_{\rm L}^2} \, 
                   \frac{ k_\perp ^2 +M^2 }{ x(1-x)}   
       \right\}
           \frac{ k_\perp ^2 +M^2 }{ x(1-x) }
            \phi^{}_{\rm L} (x, k_\perp  )          \nonumber  \\
  & & - \frac{ 2g^{}_{\rm V} }{(2\pi)^3}
         \frac{ k_\perp ^2 +M^2 }{ x (1-x) }     
  \int_0^{1} d y  \int d^2 \l_\perp \,
        \left\{  y  (1-y) +  
               \frac{   \l_\perp^2 + M^2  }{m_{\rm L}^2}\, 
        \right\} \frac{\phi^{}_{\rm L} (y, \l_\perp )}{y(1-y)}  
            \label{The_first_eq}  \\
  & & - \frac{ 2g^{}_{\rm V} }{(2\pi)^3}  
        \int_0^{1} d y  \int d^2 \l_\perp \,
          \left\{ ( \l_\perp^2 + M^2 ) +  m_{\rm L}^2 y (1-y) \right\}
           \cdot \frac{ \l_\perp ^2 +M^2 }{ m_{\rm L}^2 y (1-y)}
           \cdot
           \frac{ \phi^{}_{\rm L} (y , \l_\perp)}{y(1-y)}  \nonumber  \\
  & & + 4 \, \frac{ 2g_{\rm V}^2  }{(2\pi)^3}
        \left(\int_0^{1} d x'  \int d^2 {b}_\perp \, 
        \frac{ b_{\perp}^2 + M^2 }{ x' } \right)   \frac{1}{(2\pi)^3} 
          \int_0^{1} d y  \int d^2 \l_\perp \,
           \left\{ y  (1-y) +  
                   \frac{  \l_\perp^2 + M^2  }{m_{\rm L}^2}  \, 
           \right\}
            \frac{\phi^{}_{\rm L} (y , \l_\perp \, )}{y(1-y)}\, .\nonumber
\end{eqnarray}
Note that this result is independent of the total longitudinal momentum $P^+$.
It is not difficult to identify the origin of each term. 
On the right-hand side, the first term comes from the kinetic term 
of the Hamiltonian, the second and third terms are from $\omega_{\rm V}$,
and the last term is from $\omega_{\rm V^2}$. 

Remarkably, this complicated expression can be greatly simplified 
if one recognizes that the equation is made of three different types 
of integral. These {\it constants} are defined by the following: 
\begin{eqnarray}
  \xi & \equiv &  \int_0^{1} d y  \int d^2 \l_\perp \,
           \left\{ m_{\rm L}^2 \, y (1-y ) +  ( \l_\perp^2 + M^2 ) \, \right\}
            \frac{\phi^{}_{\rm L} (y , \l_\perp )}{y(1-y)}\, ,   \nonumber  \\
  \eta & \equiv &
        \int_0^{1} d y  \int d^2 b_\perp \,
           \frac{  b_{\perp}^2 + M^2 }{ m_{\rm L}^2  \, y (1-y ) }\, ,
       \nonumber  \\
  \nu & \equiv &
            \int_0^{1} d y  \int d^2 \l_\perp \,
           \left\{ ( \l_\perp^2 + M^2 ) +  m_{\rm L}^2 \, y (1-y)\, \right\}
           \left\{ \frac{ \l_\perp ^2 +M^2 }{ m_{\rm L}^2 \, y (1-y) } \right\}
           \frac{ \phi^{}_{\rm L} (y , \l_\perp )}{y(1-y)}\, .       \nonumber
\end{eqnarray}
In fact, these three constants are not independent of each other. 
One can verify this by first multiplying Eq.~(\ref{The_first_eq}) by 
$x (1-x) $ and then integrating it over $x$ and $k_\perp$:
\begin{equation}
  \nu = \frac{\xi}{ 1 - 2 \frac{ g^{}_{\rm V} }{(2\pi)^3  } \int_0^1 dx \int
d^2 k_\perp }
      + 2 \frac{ g^{}_{\rm V}}{(2\pi)^3  } \, \xi \, \eta\, .
\end{equation}
This relation helps to simplify the third term on the right-hand 
side of Eq.~(\ref{The_first_eq}). 
Indeed, substituting this into the third term, one gets
\begin{eqnarray}
  & & \hspace*{-1cm} \left\{ m_{\rm L}^2  + 
    \frac{ k_\perp^2+M^2 }{x(1-x)} \right\}  \, 
        \phi^{}_{\rm L} (x, k_\perp )   \nonumber  \\
  &=&  \left\{ 1 - 2 \frac{ g^{}_{\rm V}}{(2\pi)^3  } 
       \int_0^1 dz \int d^2 q_\perp \right\}^{-1}   \nonumber  \\
  & & \hskip0.3cm  \times \left[ -2 \,  \frac{ g^{}_{\rm V}}{(2\pi)^3  } \xi 
      -  \frac{g^{}_{\rm V}}{(2\pi)^3  } \, \xi \, 
         \frac{4 \,(k_\perp^2+M^2)}{m_{\rm L}^2 \, x(1-x)-(k_\perp^2+M^2)}
     \left\{ 1 - \frac{ g^{}_{\rm V}}{(2\pi)^3  } \int_0^1 dx \int d^2 k_\perp 
     \right\} \right].     \nonumber
\end{eqnarray}
Plugging the definition of $\xi$ back to this equation, one finally obtains
\begin{eqnarray}
 & & \hspace*{-0.5cm} \left\{m_{\rm L}^2 +\frac{k_\perp^2+M^2}{x(1-x)} \right\}
        \phi^{}_{\rm L} (x, k_\perp )   \nonumber  \\
  &=&  -\frac{ g^{}_{\rm V}}{(2\pi)^3  }\, 
      \left[\, 2 + \frac{ 4 \,( k_\perp^2+M^2 )}
                      {  m_{\rm L}^2  \, x(1-x) -( k_\perp^2+M^2 ) }
          \left\{ 1 - \frac{ g^{}_{\rm V}}{(2\pi)^3  } \int_0^1 dx \int d^2
k_\perp \right\} \right]     \nonumber  \\
  & &  \times\left\{ 1 - 2 \frac{ g^{}_{\rm V}}{(2\pi)^3  } 
               \int_0^1 dz \int d^2 q_\perp \right\}^{-1}    
    \int_0^{1} dy  \int d^2 \l_\perp \,
           \left\{ m_{\rm L}^2  \, y(1-y) +  ( \l_\perp^2 + M^2 ) \, \right\}
            \frac{\phi^{}_{\rm L} (y , \l_\perp )}{y(1-y)}\, . \ \    
  \label{exact_result_L}
\end{eqnarray}
This gives the potential (\ref{EQ:STLONG}) (notice the plus sign of the 
second term on 
the left-hand side while the potential is defined with minus sign, see
(\ref{LFBSE_general})). 
This is still complicated and 
we will approximate this equation by carefully taking the leading contribution
with respect to $g^{}_{\rm V}$. 

Let us integrate the above equation over the external variables $x,\, k_\perp$.
Then, we obtain a simple integral equation:
$$
  1 =  -  \frac{ g^{}_{\rm V} }{(2\pi)^3}
    \int_0^1 dx  \int d^2 k_\perp \,
 \frac{ 4 \, ( k_\perp^2 + M^2 ) }{ m_{\rm L}^2  \, x(1-x) -(k_\perp^2+M^2 ) }
     \left\{ 1 - \frac{ g^{}_{\rm V}}{(2\pi)^3  } \int_0^1 dz \int d^2 q_\perp
\right\} \, .
$$
This is actually an eigenvalue equation for $m_{\rm L}$, and we shall
treat this equation by taking the lowest order contribution with respect 
to $g^{}_{\rm V}$. Namely, we treat 
$$
  1 =  -  \frac{ g^{}_{\rm V} }{(2\pi)^3}
    \int_0^1 dx  \int d^2 k_\perp \,
  \frac{ 4 \, ( k_\perp^2 + M^2 ) }{ m_{\rm L}^2  \, x(1-x) -(k_\perp^2+M^2 )}. $$
In the bound-state equation (\ref{exact_result_L}), this approximation
corresponds to ignoring only the $g^{}_{\rm V}$ dependent term in the square
brackets. Then, one obtains 
\begin{eqnarray}
& & \hspace*{-1cm}\left\{ m_{\rm L}^2 - \frac{k_\perp^2+M^2}{x(1-x)} \right\} 
    \phi^{}_{\rm L} (x, k_\perp )   \nonumber  \\
  &=&  \left\{ 1 - 2 \frac{ g^{}_{\rm V}}{(2\pi)^3  } 
       \int_0^1 dz \int d^2 q_\perp \right\}^{-1}  
        (- 2) \, \frac{ g^{}_{\rm V}}{(2\pi)^3  } \,
           \int_0^{1} dy  \int d^2 \l_\perp \,
       \left\{ m_{\rm L}^2  +  \frac{ \l_\perp^2 + M^2 }{y(1-y)} \, \right\}
            \phi^{}_{\rm L} (y , \l_\perp).  \ \ 
  \label{eigen_L2}
\end{eqnarray}
Integrating this equation again over $x,\ k_\perp$, one finds 
$$
\int dx\, d^2\kk \left\{m_{\rm L}^2 - \frac{k_\perp^2 + M^2}{x(1-x)} \right\} 
 \phi^{}_{\rm L}(x,k_\perp) = -4 \frac{g^{}_{\rm V}}{(2\pi)^3}
\left(\int dx\, d^2\kk\right) 
\int dy\, d^2 \l_\perp \frac{\l_\perp^2+M^2}{y(1-y)}\, 
\phi^{}_{\rm L}(y, \l_\perp).
$$
Using this relation on the right-hand side of Eq.~(\ref{eigen_L2})
finally yields the bound-state equation (\ref{EQ:LONG}).
$$
 \left\{ m_{\rm L}^2 - \frac{k_\perp^2+M^2}{x(1-x)} \right\} 
    \phi^{}_{\rm L} (x, k_\perp )  =
        -  \, \frac{ g^{}_{\rm V}}{(2\pi)^3  } \,
           \int_0^{1} dy  \int d^2 \l_\perp \,
       \left\{  \frac{ 4(\l_\perp^2 + M^2) }{y(1-y)} \, \right\}
            \phi^{}_{\rm L} (y , \l_\perp).  
$$

\end{document}